\documentclass{aa}
\usepackage{graphicx}
\usepackage[varg]{txfonts}
\usepackage{longtable}
\usepackage{lscape}
\usepackage{natbib}
\bibpunct{(}{)}{;}{a}{}{,} 
\newcommand\kms{{\rm\,km\,s^{-1}}}

\newcommand\teff{T_{\rm eff}}

\begin{document}

\title{
Chemical evolution of the Galactic bulge as traced 
by \\microlensed dwarf and subgiant stars.%
\thanks{Based on data collected 
with the 6.5\,m Magellan Clay telescope at the Las Campanas 
Observatory, Chile.}
}
\subtitle{Detailed abundance analysis of OGLE-2008-BLG-209S}
\titlerunning{Chemical evolution of the Galactic bulge as traced by dwarf and
subgiant stars}

\author{
T. Bensby\inst{1}
\and
J.A. Johnson\inst{2}
\and
J. Cohen\inst{3}
\and
S. Feltzing\inst{4}
\and
A. Udalski\inst{5}
\and
A. Gould\inst{2}
\and\\
W. Huang\inst{6}
\and
I. Thompson\inst{7}
\and
J. Simmerer\inst{4}
\and
D. Ad\'en\inst{4}
 }
\institute{European Southern Observatory, Alonso de Cordova 3107, Vitacura, 
Casilla 19001, Santiago 19, Chile\\ \email{tbensby@eso.org}
\and
Department of Astronomy, Ohio State University, 140 W. 18th Avenue, 
Columbus, OH 43210, USA\\ \email{jaj,\,gould@astronomy.ohio-state.edu}
\and
Palomar Observatory, Mail Stop 105-24, California Institute of Technology, 
Pasadena, CA 91125, USA\\ \email{jlc@astro.caltech.edu}
\and
Lund Observatory, Box 43, SE-221\,00 Lund, Sweden\\
\email{sofia,\,daniela,\,jennifer@astro.lu.se}
\and
Warsaw University Observatory, A1. Ujazdowskie 4, 00-478, Warszawa, Poland\\
\email{udalski@astrouw.edu.pl}
\and
Department of Astronomy, University of Washington, Box 351580, Seattle, WA 98195, USA\\
\email{hwenjin@astro.washington.edu}
\and
The Observatories of the Carnegie Institution of Washington, Pasadena, 
CA 91101, USA\\ \email{ian@ociw.edu}
}


\date{Received 8 January 2009 / Accepted 17 March 2009}



 \abstract%
 {}
 {
 Our aims are twofold. First we aim to evaluate the robustness and 
 accuracy of stellar parameters and detailed elemental abundances that 
 can be derived from high-resolution spectroscopic observations of 
 microlensed dwarf and subgiant stars. We then aim to use microlensed 
 dwarf and subgiant stars to investigate the abundance structure and 
 chemical evolution of the 
 Milky Way Bulge. Contrary to the cool giant stars, with their extremely
 crowded spectra, the dwarf stars are hotter, their spectra are cleaner,
 and the elemental abundances of 
 the atmospheres of dwarf and subgiant stars are largely untouched 
 by the internal nuclear processes of the star. 
   }
 {
 We present a detailed elemental abundance analysis of OGLE-2008-BLG-209S, 
 the source star of a new microlensing event towards the Bulge, 
 for which we obtained a high-resolution spectrum with the MIKE 
 spectrograph on the Magellan Clay telescope. We have performed four
 different  analyses of OGLE-2008-BLG-209S. One 
 method is identical to the one used for a large comparison sample 
 of F and G dwarf stars, mainly thin and thick disc stars, in the Solar 
 neighbourhood. We have also re-analysed three previous microlensed dwarf stars 
 OGLE-2006-BLG-265S, MOA-2006-BLG-099S, and OGLE-2007-BLG-349S with the same
 method. This homogeneous data set, although small, enables a direct 
 comparison between the different stellar populations.   
 }
 {
 We find that OGLE-2008-BLG-209S is a subgiant star that has a metallicity 
 of $\rm [Fe/H]\approx -0.33$. It possesses [$\alpha$/Fe] enhancements 
 similar to what is found for Bulge giant stars at the same metallicity, and 
 what also is found for nearby thick disc stars at the same metallicity. 
 In contrast, the previous three microlensing dwarf stars have very high 
 metallicities, $\rm [Fe/H]\gtrsim +0.4$, and more solar-like abundance
 ratios, i.e. $\rm [\alpha/Fe]\approx 0$. The decrease in the [$\alpha$/Fe] 
 ratio with [Fe/H] is the typical signature of enrichment from low and 
 intermediate mass stars.  We furthermore find that the results
 for the four Bulge stars, in combination with results from studies of 
 giant stars in the Bulge, seem to favour a secular formation scenario for 
 the Bulge. 
  }
 {}
   \maketitle

\section{Introduction}

How spiral galaxies form and acquire their different stellar populations 
is largely an unsolved problem. However, since the mid-1980s the theory of 
hierarchical structure formation in a Lambda Cold Dark Matter Universe  
\citep[$\Lambda$CDM,][]{blumenthal1984} has emerged as a foundation to 
understand the properties and evolution of galaxies. In $\Lambda$CDM 
cosmological simulations, galaxies such as the Milky Way are built from 
independent smaller systems and fragments/debris from  other stellar systems 
over a time period spanning a few billion years 
\citep[e.g.][]{governato2007,read2008}. A fundamental prediction from such 
models would be that bulges formed in mergers. However, at the same time 
there is growing observational evidence that the bulges of many distant
galaxies formed by internal dynamical processes \citep[e.g.,][]{genzel2008}. 
During an early turbulent phase of a galaxy, disc 
material (gas and stars) is gravitationally driven into the central regions 
of the galaxy, building up an exponential component. This scenario
is referred to as secular evolution \citep[e.g.,][]{kormendy2004}. 

The Milky Way bulge (hereafter referred to as the Bulge, with capital B) has 
a rather broad metallicity distribution \citep[e.g.,][]{zoccali2008} and its 
metal-rich stars and globular clusters are as old as the Galactic halo
 \citep[e.g.,][]{rosenberg1999,marinfranch2008}.  This
points to a very intense star formation rate in the early history of the 
Galaxy. The Bulge thus represents an important 
link for our understanding of galactic bulges in general, and is integral to 
the question of galaxy formation and evolution \citep[e.g.,][]{wyse1997,kormendy2004}. 

Despite the fact that detailed abundance studies can provide crucial 
information on the formation and chemical enrichment history for a stellar 
population \citep[e.g.][]{mcwilliam1997}, the Bulge has long been the least 
studied stellar component of our Galaxy. This is due to the 
inherent difficulties in studying stars in the Bulge as they are distant and 
suffer from a high degree of interstellar extinction 
in the Galactic plane. However, following the pioneering study by 
\cite{mcwilliam1994}, and with the advent of 8-10 meter class telescopes, 
substantial insight into the stellar populations and the chemical history of 
the Bulge has been gained from high-resolution spectroscopic studies of bright 
K and M giant stars \citep[][]{fulbright2006,fulbright2007,cunha2006,cunha2007,
cunha2008,rich2005,rich2007,lecureur2007,zoccali2003,zoccali2008,melendez2008,ryde2009}. 

When studying the chemical history of a stellar population through detailed 
elemental abundances, one relies on the assumption that the chemical 
composition of the stars is a true measure of the elemental abundances
present in the gas from which they formed. The expected lifetimes for 
F and G dwarf stars on the main sequence, burning hydrogen to helium in their 
centres, are similar to, or possibly even longer than the current age of 
the Galaxy. 
For instance, a solar type star will spend $\sim 11$\,Gyr on the main sequence
\citep[e.g.][]{sackmann1993}
during which their atmospheres are  untouched by internal 
nuclear processes of the star \citep[e.g.][]{iben1991}. At later evolutionary 
stages, when the stars reach the red giant branch, various 
internal physical processes can erase the abundance signatures from the stellar 
atmosphere. This is the case for C, N, and Li in all giant stars, and for O, Na, 
Mg, Al in giant stars in globular clusters \citep[e.g.][]{gratton2004}.
For dwarf stars it is essentially only the rare light elements Li, Be and B 
that are depleted \citep[e.g.,][]{boesgaard2005b}, 
and as studies of chemical evolution mostly focus on heavier elements, such as 
the $\alpha$-elements, this
is one of the main reasons why observational studies of the chemical evolution 
of the Milky Way are largely based on dwarf stars \cite[e.g.][]{edvardsson1993}. 
Also, because the spectra of the intrinsically luminous metal-rich giant stars
have a very rich and often hard to identify fauna of molecular bands,
great effort has to be made to identify un-blended and weak
spectral lines \citep[see, e.g.,][]{fulbright2007}. Furthermore, for species
such as, e.g., Mg and Na, that only have a few usable lines that usually 
also are strong ($>1000$\,m{\AA}) in metal-rich dwarf stars, the lines become 
very strong in giant stars. Elemental abundance uncertainties due to NLTE effects
are also likely to be larger in giant stars than in dwarf stars 
\citep[e.g.,][]{asplund2005araa}.
Hence, there is reason to believe that the underlying assumption that the giant 
stars accurately trace the chemical evolution of a stellar population could be 
erroneous and should therefore be rigourously tested.

The concept of using microlensing to obtain spectra 
of dwarf stars in the Bulge was first demonstrated by \cite{minniti1998} 
who used Keck~I as a ``15 meter telescope" to obtain a spectrum of the 
moderately magnified ($\sim 1$\,mag) 97-BLG-45. \cite{cavallo2003} then 
presented an analysis of six microlensed stars (including the star observed 
by \citealt{minniti1998}): two cool giant stars, one subgiant star, two solar
analogues, and one of uncertain type (the one from \citealt{minniti1998}). 

After the first studies by \cite{minniti1998} and \cite{cavallo2003}, three 
microlensed Bulge dwarf stars have been observed 
\citep{johnson2007,johnson2008,cohen2008}. The results from these studies 
provide some surprising results, contradicting the results based on giant 
stars. First, it is clear that all three are unusually metal-rich. The star 
analysed by \cite{johnson2007},   at a metallicity of $\rm [Fe/H]=+0.56$, is the
most metal-rich star known to us.  As microlensing events 
have no bias with regard to the metallicity of the source star, analysing enough 
events will provide an unbiased metallicity distribution (MDF) of the Bulge. 
The three dwarf events studied in detail so far point to a much more metal-rich MDF compared 
to the one derived from
giant stars. Secondly, the stars have elemental abundance 
ratios that in general are very similar to what is found in metal-rich thin 
disc stars in the Solar neighbourhood, i.e. they do not show the high 
$\rm [\alpha/Fe]$ ratios that is found in the Bulge giant stars. Although the 
dwarf sample is still very small, these findings hint that the assumption 
that giant stars give us a complete picture of the Bulge may not be correct.

\begin{figure}
\resizebox{\hsize}{!}{
\includegraphics[bb=28 150 590 570,clip]{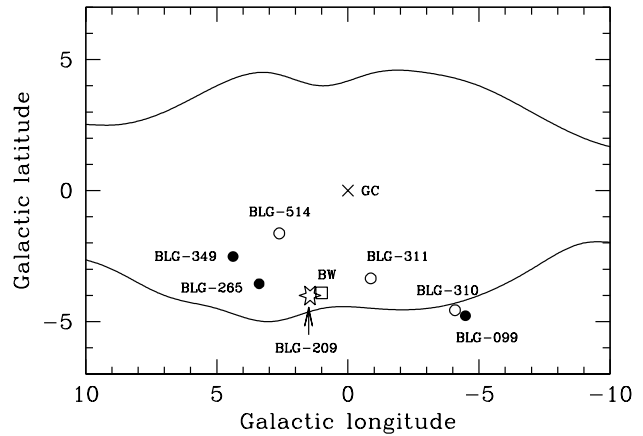}}
\caption{Dwarf stars in the Bulge for which high-resolution spectra
have been obtained. Filled circles mark OGLE-2006-BLG-265S and
MOA-2006-BLG-099S already published by \cite{johnson2007} and 
\cite{johnson2008}, respectively, and OGLE-2007-BLG-349S published
by \cite{cohen2008}. The ``star" marks OGLE-2008-BLG-209S being
analysed here for the first time, and the open circles
mark OGLE-2007-BLG-514S, MOA-2008-BLG-310S, and MOA-2008-BLG-311S
that currently are being analysed elsewhere. The curved lines
show the outline of the Bulge based on observations with the 
COBE satellite. The open square marks the position of Baade's window 
(BW), and the Galactic centre (GC) is also marked.
        \label{fig:allevents}}
\end{figure}

In this paper we present a detailed abundance analysis of
OGLE-2008-BLG-209S, the source star of a microlensing event towards 
the Bulge. We have performed four different analyses of 
the star, enabling a comparison of the robustness of the derived stellar 
parameters and elemental abundances. 
The methods we use have all been applied either in previous studies of of lensed Bulge
dwarf stars or in large studies of nearby dwarf stars.
One method utilises the spectral 
linelist, atomic data, model stellar atmospheres, and method to find the 
stellar parameters that are all very similar to what is used 
in the studies of nearby F and G dwarf stars
\citep[][and Bensby et al.~2009, in prep.]{bensby2003,bensby2005}.
This enables direct comparisons between the Bulge stars and the local 
disc stars. The other two methods are similar to the ones used for
the analyses of OGLE-2006-BLG-265S, MOA-2006-BLG-099S, and OGLE-2007-BLG-349S
in \citep{johnson2007,johnson2008,cohen2008}. Spectra for these three other 
microlensed Bulge dwarf stars were made available for this study and we have 
re-analysed them with the method akin to the one used in 
\cite{bensby2003,bensby2005} and Bensby et al.~(2009, in prep.).

\section{Observations and data reduction}
\label{sec:observations}

In Fig.~\ref{fig:allevents} we show the positions of the microlensed
dwarf stars in the Bulge that have been observed with high-resolution 
spectrographs so far. In this figure, three additional events are marked.
These three are  OGLE-2007-BLG-514S, MOA-2008-BLG-310S, and 
MOA-2008-BLG-311S. They are currently being analysed and will be 
published elsewhere (Epstein et al., in prep.; Cohen et al., in prep.).

\subsection{Nomenclature}

First we want to clarify the nomenclature used to describe 
microlensing events.
The actual microlensing event is given a name depending
on the project which discovered it (e.g., MOA or OGLE), year it took place, 
the direction in which the event occurred (in our case BLG for the
Bulge), and a running number, e.g.,
OGLE-2008-BLG-209. When referring to the source star of the
microlensing event an "S" is added to the event name, i.e.,
OGLE-2008-BLG-209S. Other extensions are given for the lens and 
for any planets orbiting the lens. 

\subsection{OGLE-2008-BLG-209S}

\begin{figure}
\resizebox{\hsize}{!}{
\includegraphics[bb=20 250 550 750,clip]{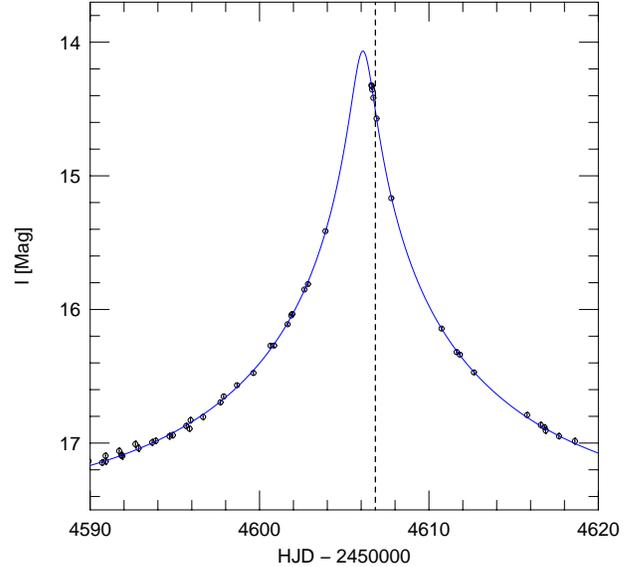}}
\caption{OGLE photometry of the microlensing event. Maximum brightness
         was estimated to occur on  HJD2454606.097\,($\pm0.004$) 
         i.e., UT2008-05-20.06. The solid line is a 
         theoretical fit to the microlensing event. 
         The vertical dotted line indicates
         when observations (start of first exposure) were carried out 
         with MIKE (HJD2454606.838).
         \label{fig:event}}
\end{figure}

In May 2008 the OGLE early warning 
system\footnote{\tt http://ogle.astrouw.edu.pl/ogle3/ews/ews.html} 
\citep{udalski2003} put out an alert that a new microlensing event, 
OGLE-2008-BLG-209, was developing and would reach maximum 
brightness on HJD2454606.097. The faintness of the source star before 
the microlensing event made 
it likely for the source star to be a dwarf star in the Bulge, located 
close to Baade's window at $(l,\,b)=(1.4,\,-4.0)$\,deg 
(see Fig.~\ref{fig:allevents}). Fortunately we had observing time on 
Magellan\,II telescope at this time allowing us to obtain a 
high-resolution spectrum 
of the object with the MIKE spectrograph \citep{bernstein2003}. At the 
time of observation on May 20, 2008 (HJD2454606.838) the event had just 
passed maximum brightness and had a magnitude of $I\sim 14.5$, about 
20 times brighter (approximately 3 magnitudes) than before being 
microlensed (see Fig.~\ref{fig:event}). Three 1800\,s exposures were 
obtained using a slit width of 0.7\,arcsec, resulting in a spectrum, 
with continuous wavelength coverage from  3200\,{\AA} to 10\,000\,{\AA},
recorded on two CCDs. With a 0.7\,arcsec slit the  
resolving power is $R\sim 55\,000$ on the blue CCD and $R\sim 47\,000$ 
on the red CCD. The data were reduced with the Carnegie Observatories 
MIKE Python pipeline (D. Kelson, private communication) and the final 
spectrum used in the analysis consists of the three individual spectra 
co-added and then divided by the blaze function. Typical signal-to-noise 
ratios are $S/N\sim20$ per pixel in the spectrum from the blue CCD and $S/N\sim 30$ 
per pixel in the spectrum from the red CCD (as measured in the 
continuum of regions free of spectral lines).

\subsection{OGLE-2006-BLG-265S, MOA-2006-BLG-099S, and OGLE-2007-BLG-349S}

\cite{johnson2007} presented a detailed abundance analysis of 
OGLE-2006-BLG-265S. At the time 
of observation it was magnified by a factor of $\sim 145$ and, 
by using the HIRES spectrograph on the Keck telescope, they obtained a 
spectrum of high-resolution ($R\sim45\,000$) and relatively high 
signal-to-noise ratio ($S/N\sim 60$ per resolution element).  

A year later, \cite{johnson2008} presented a detailed abundance analysis
of MOA-2006-BLG-099S, another microlensed Bulge dwarf star. This object 
was at the time of observation magnified by a factor of $\sim110$, 
and a spectrum was obtained with the MIKE spectrograph on the 
Magellan\,II telescope. Compared to OGLE-2006-BLG-265S, the spectrum of 
MOA-2006-BLG-099S is of lower quality with a resolution of only 
$R\sim 19\,000$ red-ward of 4800\,{\AA} and $S/N\sim 30$ per pixel.

OGLE-2007-BLG-349S, the third dwarf reanalysed here, was observed 
on September 5, 2007 at the Keck Observatory. Three consecutive spectra, 
each 1350\,s in length, were obtained with HIRES-R at the Keck~I 
telescope in a configuration with coverage from 3900 to 8350\,{\AA}, 
with small gaps between the orders beyond 6650\,{\AA}. The slit was 
0.86\,arcsec wide, giving a spectral resolution of $R\sim48,000$.
The magnifications at the time of the recording of the three spectra were 
350, 390, and 450. The signal-to-noise ratio of the resulting summed spectrum is
$S/N\sim30$ per spectral resolution element for wavelengths below 
5000\,{\AA} and $S/N\sim90$ for 
wavelengths above 5500\,{\AA}.

Further details about observations and data reductions for these three 
stars can be found in \cite{johnson2007,johnson2008} and \cite{cohen2008}.

\begin{figure*}
\centering
\resizebox{17cm}{!}{
\includegraphics[angle=-90,bb=60 28 540 810,clip]{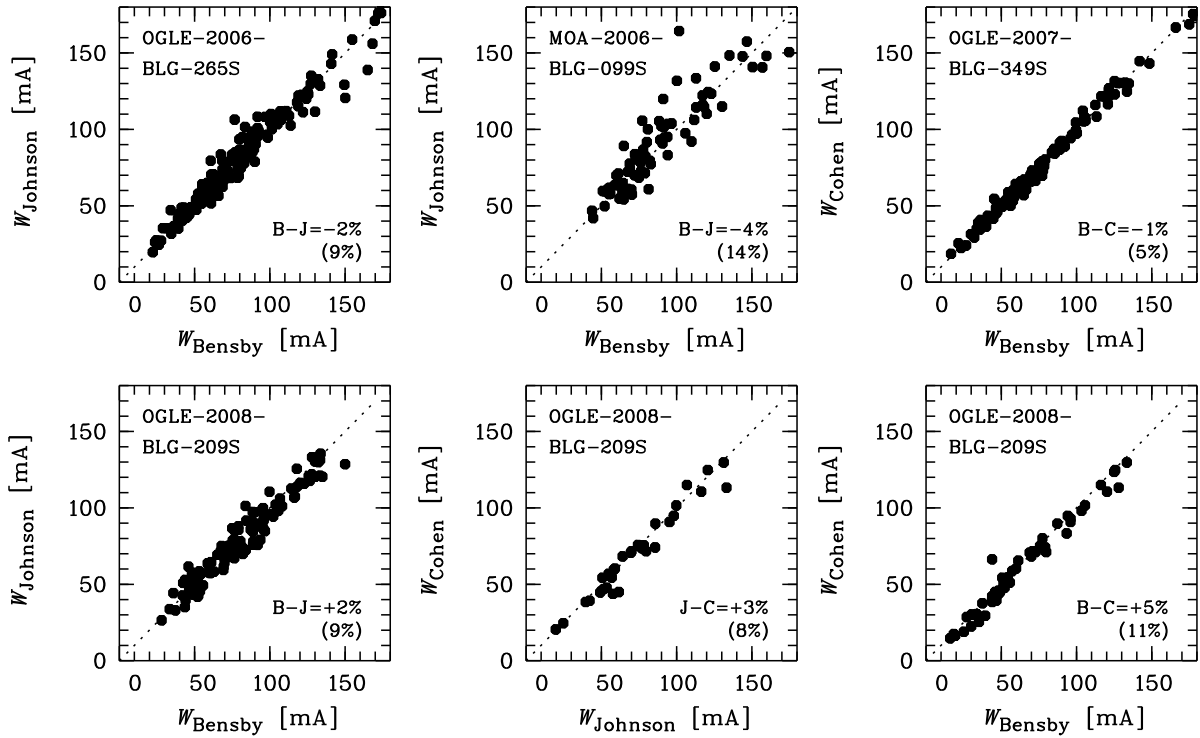}}
\caption{
         Comparison of equivalent widths measured independently
         by Bensby (B), Cohen (C) and Johnson (J) for the four
         stars that have been analysed. Differences and standard
         deviations (in parentheses) are indicated in the figure.
         The dotted line in each plot indicates the one-to-one relation.
\label{fig:eqws}}
\end{figure*}

\subsection{Radial velocities}
\label{sec:rv}

For OGLE-2008-BLG-209S we measure a velocity shift of 
$-190.1\pm0.4\,\kms$ in the spectrum from the blue CCD and 
$-189.0\pm0.1\,\kms$ in the spectrum from the red CCD. As the red 
spectrum is of higher quality we use this one only for the radial 
velocity. Adding the heliocentric correction, which at the time of 
observation was $+15.4\,\kms$, the heliocentric radial velocity of 
OGLE-2008-BLG-209S is $-173.6\,\kms$. The three 
stars, OGLE-2006-BLG-265S, MOA-2006-BLG-099S, and OGLE-2007-BLG-349S
have  radial velocities of $+99$, $-154$, and $+113\,\kms$, 
respectively \citep{johnson2007,johnson2008,cohen2008}. 

\section{Abundance analysis}
\label{sec:analysis}

Stellar parameters and elemental abundances will be determined 
independently by three of us (T.B., J.J., and J.C.) using different 
approaches. The different methods will use their own linelists,
atomic data, and model stellar atmospheres as used in the various
previous studies of Bulge as well as local dwarf stars. 
As equivalent widths are the fundamental ingredient 
for all four methods, we will start by checking the agreement between 
the different measurements for lines that are in common.

\subsection{Linelists and equivalent width comparisons}
\label{sec:eqws}

The analysis by Bensby (methods 1 and 2 below) uses the same
spectral line list and atomic data as in 
\citep{bensby2003,bensby2005}. However, it has now been expanded by
another 50 \ion{Fe}{i} lines (see Bensby et al.~2009, in prep.;
and Table~\ref{tab:onlinebensby}).
Equivalent widths were measured by hand using the 
IRAF\footnote{IRAF is distributed by the National Optical Astronomy 
Observatory, which is operated by the Association of Universities for 
Research in Astronomy, Inc., under co-operative agreement with the 
National Science Foundation.} task {\sc splot}. Gaussian line profiles 
were fitted to the observed lines and in special cases of strong Mg, 
Ca, Si, and Ba lines, we used Voigt profiles instead to better account 
for the extended wing profiles of these lines.

The analysis by Johnson (method 3 below) uses a linelist that for most 
elements was taken from  \cite{bensby2003,bensby2004}. The equivalent
widths were measured using 
SPECTRE\footnote{\tt http://verdi.as.utexas.edu/spectre.html} 
(C. Sneden, 2007, private communication). For Ba, synthetic spectra were
compared to the observed spectra to determine the abundance.
The effect of hyperfine splitting (HFS) in the Ba lines was included 
and the Ba HFS constants and $\log gf$-values 
were taken from the sources listed in \cite{johnson2006}. 

The analysis by Cohen (method 4 below) uses the linelist given 
in \cite{cohen2008}. Equivalent widths were measured using an automatic 
Gaussian fitting routine, after which stronger lines were checked by 
hand in order to make sure that damping wings were picked up when 
appropriate. Elements with only a few detected lines were also checked 
by hand. 

\begin{table*}
\centering
\caption{
Comparison of equivalent widths measured by Bensby, Johnson and Cohen.
For each element we give the total number of lines that each
of the different linelists contain, the average differences (in percent),
the 1-sigma spreads, and the number of lines in common (in parentheses). 
\label{tab:ews}
}
\setlength{\tabcolsep}{1.5mm}
\tiny
\begin{tabular}{l|rrrlll|rrl|rrl|rrl}
\hline\hline
\noalign{\smallskip}
       &
       \multicolumn{6}{c|}{OGLE-2008-BLG-209S}  &
       \multicolumn{3}{c|}{MOA-2006-BLG-099S}  &
       \multicolumn{3}{c|}{OGLE-2006-BLG-265S}   &
       \multicolumn{3}{c}{OGLE-2007-BLG-349S}  \\ 
       &
       Ben  &
       John &
       Coh  &
       \multicolumn{1}{c}{Ben-John} &
       \multicolumn{1}{c}{Ben-Coh}  &
       \multicolumn{1}{c|}{John-Coh} &
       Ben  &
       John &
       \multicolumn{1}{c|}{Ben-John} &
       Ben  &
       John &
       \multicolumn{1}{c|}{Ben-John}  &
       Ben  &
       Coh &
       \multicolumn{1}{c}{Ben-Coh}   \\
\noalign{\smallskip}
                &
            $N$ &
            $N$ &
            $N$ &
            \multicolumn{1}{c}{\% $\pm\sigma$ ($N$) } &
            \multicolumn{1}{c}{\% $\pm\sigma$ ($N$) } &
            \multicolumn{1}{c|}{\% $\pm\sigma$ ($N$) } &
            $N$ &
            $N$ &
            \multicolumn{1}{c|}{\% $\pm\sigma$ ($N$) } &
            $N$ &
            $N$ &
            \multicolumn{1}{c|}{\% $\pm\sigma$ ($N$) } &
            $N$ &
            $N$ &
            \multicolumn{1}{c}{\% $\pm\sigma$ ($N$) } \\
\noalign{\smallskip}
\hline
\noalign{\smallskip}
Total        & 284 & 146 & 109 & $+2\pm9$ (135) & $+5\pm11$ (55) & $+3\pm8$ (32) & 251 &105 & $-4\pm14$ (76) & 188 & 219 & $-2\pm8$ (172) & 181 & 249 & $-1\pm5$ (117)\\
\ion{Fe}{i}  & 145 &  72 & 101 & $-1\pm10$ (61) & $+4\pm11$ (49) & $+3\pm8$ (31) & 122 & 35 & $-5\pm11$ (19) & 91  &  98 & $-2\pm8$ (79)  &  91 & 135 & $-0\pm3$ (59) \\ 
\ion{Fe}{ii} &  19 &   2 &   8 & $+4\pm8$  (2)  & $+12\pm12$ (6) & $+9\pm0$ (1)  &  14 &  4 & $-3\pm9$ (3)   & 21  &   9 & $-3\pm14$ (8)  &  14 &  11 & $-2\pm8$ (11) \\
\noalign{\smallskip}
\hline
\end{tabular}
\end{table*}

Figure~\ref{fig:eqws} shows comparisons of the equivalent widths as 
measured by Bensby, Johnson, and Cohen. There is generally a good agreement. 
For OGLE-2008-BLG-209S Bensby's measurements are on average 5\,\% larger
than Cohen's (55 lines in common), and 2\,\% larger than Johnson's 
(135 lines in common), which in turn are 3\,\% larger than Cohen's 
(34 lines in common). For OGLE-2006-BLG-265S the Bensby measurements are
just 2\,\% smaller (172 lines in common) than the measurements by 
\cite{johnson2007}, and for MOA-2006-BLG-099S the Bensby measurements are
4\,\% smaller (76 lines in common) than the ones by \cite{johnson2008}.
For OGLE-2007-BLG-349S, which has the highest SNR spectrum of the four 
microlensed stars, the agreement is even better, the Bensby 
measurements are only 1\,\% smaller  than the 
measurements by \cite{cohen2008} (117 lines in common). 
These differences relate to {\sl all} 
equivalent widths from all elements and ions measured. By comparing the 
equivalent widths of the \ion{Fe}{i} and \ion{Fe}{ii} lines only, we see 
that there are significant larger average differences for the \ion{Fe}{ii} 
lines than for the bulk of lines (see Table~\ref{tab:ews}). For instance, 
for OGLE-2008-BLG-209S the differences in equivalent widths between Bensby's 
and Cohen's measurements of \ion{Fe}{ii} lines reach 12\,\%. How these 
differences affect the stellar parameters is further investigated
in Sect.~\ref{sec:results209}.

\begin{figure}
\centering
\resizebox{\hsize}{!}{
\includegraphics[angle=-90,bb=20 28 570 600,clip]{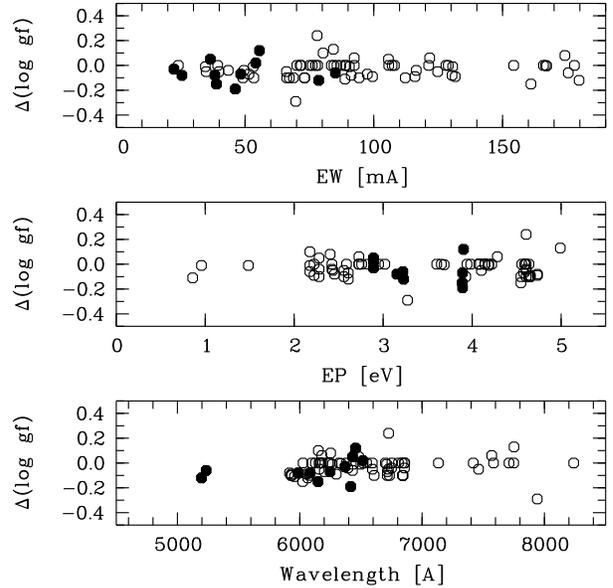}}
\caption{Comparison of \ion{Fe}{i} and \ion{Fe}{ii} oscillator strengths 
($\log gf$) used in methods 1 and 2 
\citep[values taken from][and Bensby et al.~2009, in prep.]{bensby2003} 
and method 4 \citep[values taken from][]{cohen2008}. The plots show
Bensby's values minus Cohen's values as a function of line strength (top panel), 
in this case Cohen's measurements for OGLE-2007-BLG-349S; 
lower excitation potential (middle panel); and 
wavelength (bottom panel). Open and filled circles mark \ion{Fe}{i} and
\ion{Fe}{ii} lines, respectively. 
On average the Cohen values are $0.03\pm0.07$\,dex
larger for the \ion{Fe}{i} lines and 
and $0.05\pm0.09$\,dex larger for the \ion{Fe}{ii} lines (i.e. giving lower abundances).
\label{fig:comp_loggf}}
\end{figure}

Figure~\ref{fig:comp_loggf} shows a comparison of oscillator strengths
($\log gf$-values) for \ion{Fe}{i} and \ion{Fe}{ii} lines.
Since  the Johnson studies largely have adopted the linelists and atomic 
data as  given in \cite{bensby2003} we only make comparisons between 
the Bensby and Cohen linelists. The figure shows the differences in the 
$\log gf$-values as a function of wavelength,  as a function of the 
lower excitation potential of the line, and as a function of the equivalent 
width. From these plots we see that there are differences present, however 
with no obvious trends with either excitation potential, line strength, 
or wavelength. Average differences are $-0.03\pm0.07$\,dex for \ion{Fe}{i} lines 
and $-0.05\pm0.09$\,dex for \ion{Fe}{ii} lines, with the values from Cohen
linelist being the larger ones. This means that for a given combination 
of stellar parameters, the average {\sl absolute} iron abundances by 
\cite{cohen2008} will come out $0.03$\,dex lower if based on \ion{Fe}{i} 
lines and $0.05$\,dex lower if based on \ion{Fe}{ii} lines, if the same
equivalent widths and model atmospheres are used. 

As the tuning of the stellar parameters are based on absolute
abundances from \ion{Fe}{i} and \ion{Fe}{ii} lines (i.e. not normalised 
to the solar abundances) these differences may have an effect on how 
the stellar parameters come out from the different methods. This will 
be investigated further in Sect.~\ref{sec:results209}. Note that it is 
essentially only for the Fe lines that good accuracy of the $\log gf$-values 
are important. Abundances based on lines from other elements will be 
normalised to those of the Sun, cancelling, to first order, out any 
uncertainties in the $\log gf$-values. 

Also, the two \ion{Fe}{i} lines at 6726\,{\AA} and 7941\,{\AA} clearly show larger deviations
than the other lines (see Fig.~\ref{fig:comp_loggf}). 
In Table~\ref{tab:onlinebensby}, where we give abundances
for all lines we see that the solar abundance from the $\lambda$6726 line
is too low, and from the $\lambda$7941 line it is too high. The normalised 
abundances from these two lines for the four stars agree however well
with the normalised abundances from the other \ion{Fe}{i} lines. So, it seems that
the $\log gf$ values for these two lines could be significantly off. However,
as disregarding the two lines have no effects on the stellar parameters nor the
final abundances, we will, for the moment, keep them as they are. 

\begin{figure*}
\resizebox{\hsize}{!}{
\includegraphics[angle=-90,bb=60 28 570 450,clip]{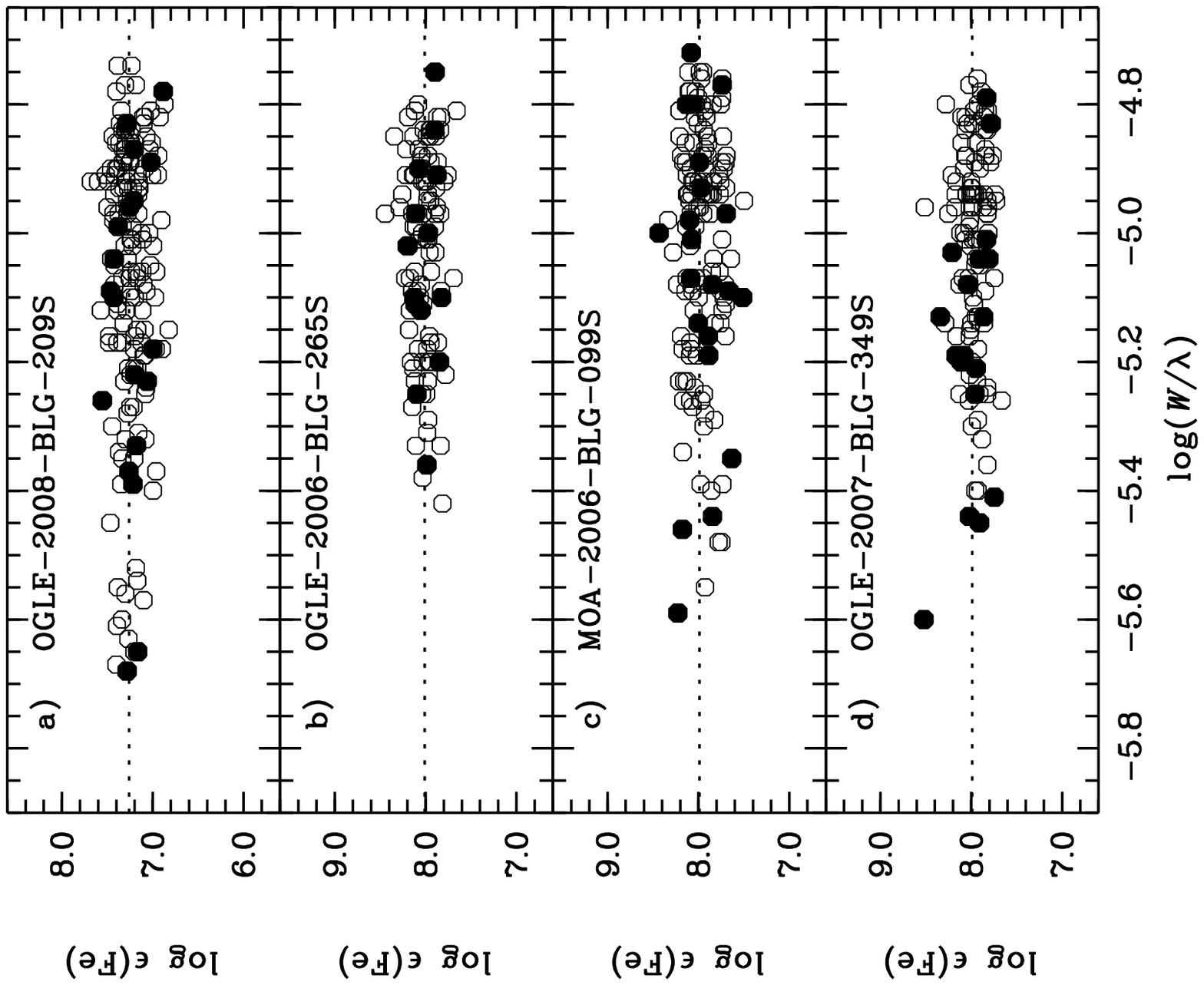}
\includegraphics[angle=-90,bb=60 28 570 450,clip]{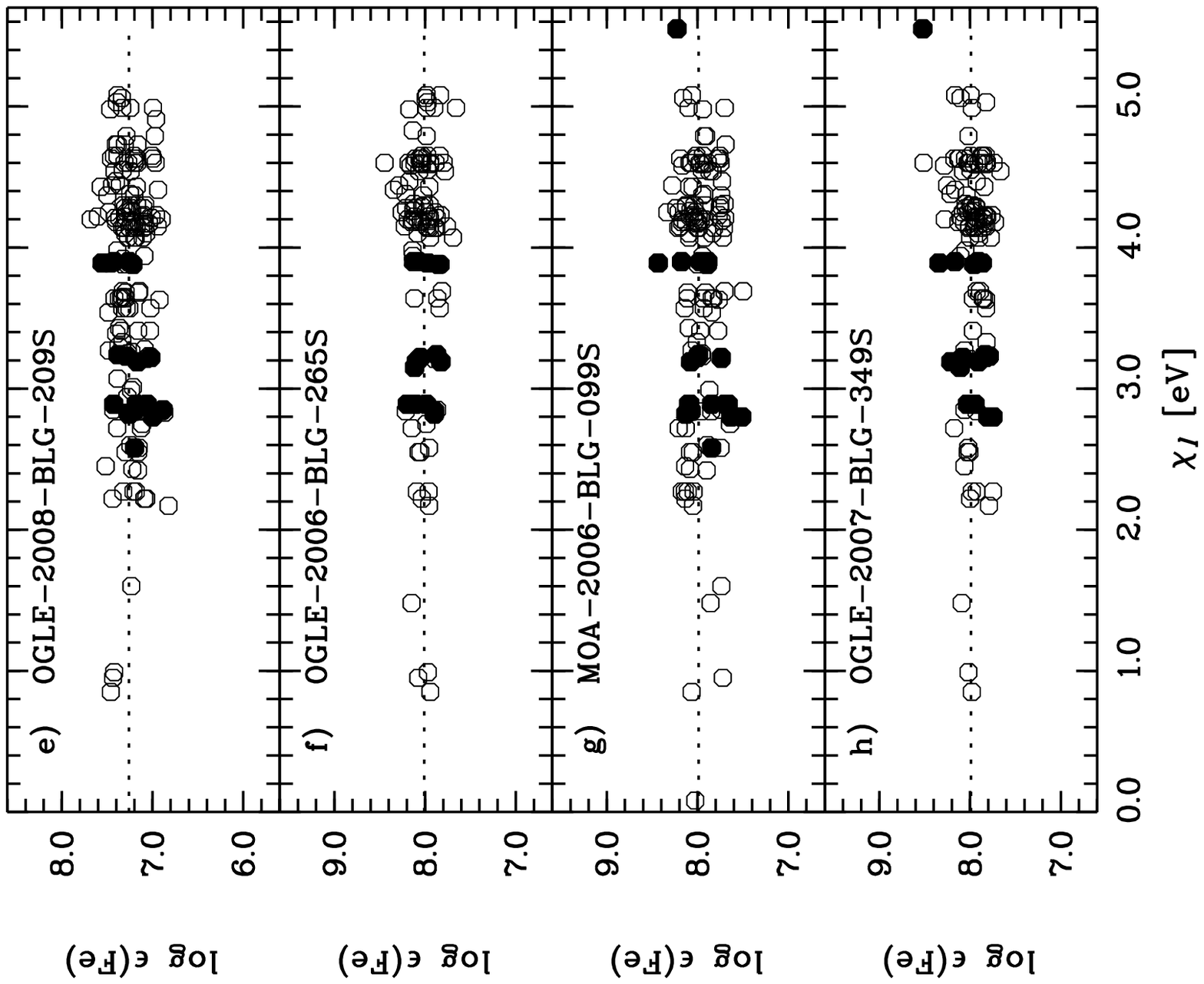}}
\caption{Absolute Fe abundances versus reduced line strength and 
         versus lower excitation potential of the line. Empty circles
         are abundances from Fe\,{\sc i} lines and filled circles
         from Fe\,{\sc ii} lines. The dashed lines are the linear 
         regression to the Fe\,{\sc i} abundances.
        \label{fig:fetrends}}
\end{figure*}

The Bensby linelist, with measured equivalent widths and calculated 
elemental abundances for OGLE-2008-BLG-209S, OGLE-2006-BLG-265S,
MOA-2006-BLG-099S, and OGLE-2007-BLG-349S is given in 
Table~\ref{tab:onlinebensby}.

\subsection{Method 1: Bensby analysis with gravity from ionisation balance}
\label{sec:method1}

This is a standard method based one-dimensional plane-parallel LTE 
model stellar atmospheres that were calculated with the Uppsala MARCS 
code \citep{gustafsson1975,edvardsson1993,asplund1997}. 
Elemental abundances were calculated with the associated suite
of programs for abundance analysis and line synthesis 
(EQWIDTH and SPECTRUM, Edvardsson, Eriksson, \& Gustafsson, 2000, private comm.).
The resulting 
stellar parameters and elemental abundances are all based on the 
equivalent widths measured by T.~Bensby. The choice of these MARCS model 
atmospheres ensures that the results based on this method are fully 
compatible with our studies of nearby dwarf stars belonging to the 
Galactic thin and thick discs 
\citep[][and Bensby et al.~2009, in prep.]{bensby2003,bensby2005}.
To find the stellar parameters we use a grid of approximately 
$12\,000$ MARCS model stellar atmospheres spanning metallicities between 
$\rm -2.2\leq[Fe/H]\leq+0.7$ in steps of 0.1\,dex; surface gravities 
between $3.2\leq\log g\leq4.8$ in steps of 0.1\,dex, and temperatures 
between $4500\leq T_{\rm eff} \leq 7000$\,K in steps of 100\,K  
(see Bensby et al.~2009, in prep.). 
The models have chemical compositions scaled relative to the
standard solar abundances of \cite{asplundgrevessesauval2005}.
However, to better reflect the actual composition of the stars, 
starting at solar metallicity, the models have $\alpha$-enhancements
that increase with decreasing metallicity, reaching $[\rm \alpha/Fe]=0.40$
at $\rm [Fe/H]=-1.0$. Below $\rm [Fe/H]=-1.0$ the $\alpha$-enhancement is 
constant at $[\rm \alpha/Fe]=0.40$, and above the solar metallicity, 
$\rm [Fe/H]=0$, there is no $\alpha$-enhancement.
When calculating abundances the broadening by collisions were taken 
from \cite{anstee,barklem1,barklem3,barklem2,barklem4}. For lines 
not included in these works the classical Uns\"old broadening was
used \citep[see][]{bensby2003}.

The stellar parameters are determined using the abundances from 
\ion{Fe}{i} and \ion{Fe}{ii} lines. The basic concepts of the method 
are as follows:
(1) The effective temperature ($\teff$) is determined by requiring
zero slope in the diagram where abundances from the \ion{Fe}{i}
lines are plotted versus
the lower excitation energy of the line ($\chi_{\rm l}$), i.e. 
excitation equilibrium; 
(2) The microturbulence parameter ($\xi_{\rm t}$) is determined by 
requiring 
zero slope in the diagram where abundances from \ion{Fe}{i} lines
are plotted versus
the measured line strength ($\log (W/\lambda))$; 
(3) The surface gravity is determined by requiring that 
the average abundance from the
\ion{Fe}{i} lines equals the average abundance
from \ion{Fe}{ii} lines,
i.e. ionisation equilibrium. This method was used because we do not 
know the apparent magnitude nor the distance of the star.
All of these balances are coupled and it is not possible to first tune 
one parameter and then another. All parameters have to be found 
simultaneously, and after extensive testing we find that an acceptable 
solution can only be found for one set of parameters for a given star.
Final diagnostic plots for $T_{\rm eff}$ and $\xi_{\rm t}$ are shown 
in Fig.~\ref{fig:fetrends}. The ionisation balance was deemed good when 
the average output abundances from the \ion{Fe}{i} and \ion{Fe}{ii} 
lines agreed within 0.01\,dex. Also, to avoid uncertainties in the stellar 
parameters arising due to saturation 
effects in strong lines, only \ion{Fe}{i} and \ion{Fe}{ii} lines with 
measured equivalent widths smaller than 90\,m{\AA} were used in the 
determination of the stellar parameters. 

There are indications that abundances from some \ion{Fe}{i} lines may 
be sensitive to departures from LTE which could invalidate our assumption of 
ionisation balance when determining the surface gravity.
However, the predicted correction for a dwarf star at $\rm [Fe/H]=-0.5$ is 
small, around 0.05\,dex \citep{thevenin1999}. Furthermore, in 
\cite{bensby2003,bensby2005} and Bensby et al.~(2009, in prep.)  we have 
analysed a total of $\sim 700$ nearby F and G dwarf stars in the Solar 
neighbourhood, all having accurate distances from the new reduction by 
\cite{vanleeuwen2007} of the parallaxes from the {\sl Hipparcos} satellite.
The fact that for these $\sim700$ F and G dwarf stars we do not require
ionisation equilibrium, but anyway find a very good agreement between 
abundances from the two ionisation stages of iron 
($\rm [\ion{Fe}{i}/\ion{Fe}{ii}]= 0.01\pm 0.07$), means that ionisation 
balance should be a valid option to use for F and G dwarf stars when the 
distance and/or the apparent magnitude is unknown. Hence,  Method 1 and 
the method for nearby dwarf stars as presented in \cite{bensby2003,bensby2005} and 
Bensby et al.~(2009, in prep.), where the distances and magnitudes of the 
stars are known,  should produce results that are fully compatible. 

\subsection{Method 2: Bensby analysis with gravity based on microlensing assumptions}
\label{sec:method2}

This method is similar to method 1 and only differs in the
way the surface gravity is estimated.
Instead of using ionisation balance the surface gravity will be 
determined using an absolute magnitude 
estimated from microlensing techniques. As the linelists, equivalent
width measurements, model stellar atmospheres, are the same
as in method 1, we only describe how the absolute magnitude is 
estimated.

De-reddened colours and magnitudes of the source can be estimated 
using standard microlensing techniques \citep[e.g.][]{yoo2004}. 
The method for determining the colour does not make any assumption 
about the absolute reddening, nor about the ratio of selective to 
total extinction.  It only assumes that the reddening to the 
microlensed source is the same as the reddening to the red clump, and that 
the red clump has $(V-I)_{0} = 1.00$, the same as the local 
{\sl Hipparcos} red clump. However, in principle, the Bulge red 
clump could be different due to different age and composition of the Bulge
as compared to the Solar neighbourhood stars. From the spectroscopic 
temperatures of the three previous Bulge dwarf stars 
\citep{johnson2007,johnson2008,cohen2008} it is found that the mean 
Bulge clump has $(V-I)_{0} = 1.05$. This value is in good agreement 
with a completely independent estimate based on taking into account 
the age and differences in composition between the Bulge and the 
Hipparcos stars (David Bennett, private  communication). 
$(V-I)_{0} = 1.05$ for the red clump will be the value used in 
this study. The absolute de-reddened colour is then derived from the 
colour offset between the microlensing source and the clump in the 
instrumental colour-magnitude diagram (CMD). $V$ and $I$ measurements 
give a colour estimate of $(V-I)_{0} = 0.73$ for OGLE-2008=BLG-209S.

The de-reddened apparent magnitude of the source is determined in a
similar way, using an assumed de-reddened apparent magnitude of the
red clump $(I_{0}=14.32)$ and the offset between the apparent magnitudes
of the source and the clump on the instrumental CMD.  The
instrumental magnitude of the source is determined from the
microlensing model. This yields $I_{0} = 17.05$ for OGLE-2008-BLG-209S.
The absolute magnitude of the source is then determined from
assuming that the red clump stars are at the same distance
as the source. For OGLE-2008-BLG-209S the estimate for the apparent 
magnitude is $V_{0}=17.68$, and the absolute magnitude is then 
$M_{V} = 3.26$ (calculated assuming a distance of 8\,kpc). 

Using this absolute magnitude the stellar parameters are then found in 
the same way as for method 1, except that ionisation equilibrium is 
not required. 

\begin{table*}
\centering
\caption{
\label{tab:onlinebensby}
Bensby's linelist with measured equivalent widths and calculated elemental abundances for
OGLE-2008-BLG-209S. Also included are the equivalent widths and elemental abundances
measured by Bensby in the solar spectrum and in spectra for OGLE-2006-BLG-265S, MOA-2006-BLG-099S, and
OGLE-2007-BLG-349S provided by Cohen and Johnson.
For each line we give the $\log gf$ value, lower excitation potential ($\chi_{\rm l}$), measured equivalent widths ($W_{\rm \lambda}$),
derived absolute abundance ($\log \epsilon (X)$), abundance normalised to the Sun (on a line-by-line basis) ([$X$/H]). The ``flag" for the Sun indicates that a line has not been
measured in the solar spectrum and that the abundance given (and used in the normalisation) for that
line is the average abundance based on all the other lines, that could be measured in the
solar spectrum, of the same species.
The table is only available in the online version of the paper and in electronic
form at the CDS via anonymous ftp to {\tt cdsarc.u-strasbg.fr (130.79.125.5)}
or via {\tt http://cdsweb.u-strasbg.fr/Abstract.html}.
        } 
\tiny   
\setlength{\tabcolsep}{1.0mm}
\begin{tabular}{rlcrc|rcc|rcr|rcr|rcr|rcr}
\hline\hline
\noalign{\smallskip}
        &
        &
        &
        &
        &
\multicolumn{3}{c|}{Sun}        &
\multicolumn{3}{c|}{OGLE-2008-BLG-209}  &
\multicolumn{3}{c|}{OGLE-2006-BLG-265}  &
\multicolumn{3}{c|}{MOA-2006-BLG-099}  &
\multicolumn{3}{c}{OGLE-2007-BLG-349}  \\
\noalign{\smallskip}
\multicolumn{2}{c}{Element}     &
Wavelength &
$\log gf$        &
$\chi_{\rm l}$    &
$W_{\rm \lambda,\odot}$ &
$\log \epsilon (X)_{\odot}$ &
flag    &
$W_{\rm \lambda}$ & 
$\log \epsilon (X)$ &
[$X$/H] &
$W_{\rm \lambda}$ & 
$\log \epsilon (X)$ &
[$X$/H] &
$W_{\rm \lambda}$ & 
$\log \epsilon (X)$ &
[$X$/H] &
$W_{\rm \lambda}$ &
$\log \epsilon (X)$ &
[$X$/H]         \\
\noalign{\smallskip}
\hline
\noalign{\smallskip}
 Al & 1 & 5557.063  & -2.21 & 3.14 &  10.4 & 6.44 & -- &  17.1 & 6.45 &  0.01 &  39.9 & 7.04 &  0.60 &  29.4 & 7.00 &  0.56 &  37.9 & 6.89 &  0.45   \\
 Al & 1 & 6696.023  & -1.62 & 3.14 &  44.9 & 6.62 & -- &  47.9 & 6.45 & -0.17 &  87.8 & 7.09 &  0.47 &  58.8 & 6.84 &  0.23 &  89.5 & 7.03 &  0.41   \\
\vdots &
\vdots &
\vdots &
\vdots &
\vdots &
\vdots &
\vdots &
\vdots &
\vdots &
\vdots &
\vdots &
\vdots &
\vdots &
\vdots &
\vdots &
\vdots &
\vdots &
\vdots &
\vdots &
\vdots \\
\noalign{\smallskip}
\hline
\end{tabular}
\end{table*}

\subsection{Method 3: Analysis by Johnson}

TurboSpectrum \citep{alvarez1998}, a 1-dimensional LTE code, was used 
to derive elemental abundances. The input model stellar atmospheres 
were interpolated\footnote{The interpolator by T.~Masseron, and the
new MARCS models are available 
on the MARCS website {\tt http://marcs.astro.uu.se}.} 
in a grid of the new MARCS 2008 model atmospheres \citep{gustafsson2008}.  
These models have chemical compositions scaled relative to the
standard solar abundances in \cite{grevesse2007}, but with 
$\alpha$-enhancements for sub-solar metallicities that are the
same as used for the MARCS 1997 models in methods 1 and 2 above 
(see Sect.~\ref{sec:method1}).
When calculating elemental abundances 
the treatment of damping from \cite{barklem4} was used.  

Effective temperatures and the microturbulence velocities were determined 
using standard techniques in a similar way to Method 1, including
$\log g$ that was determined by demanding ionisation equilibrium. 
The final parameters 
adopted for OGLE-2008-BLG-209S were $\teff=5250$\,K, $\log g=3.5$, 
metallicity $\rm [Fe/H]=-0.40$, and $\xi_{\rm t}=1.4\,\kms$. 

Further details of the method can be found in \cite{johnson2007,johnson2008}. 
It should be noted that an attempt to measure the effective temperature 
from the Balmer H$\alpha$ line, as was done for MOA-2006-BLG-099S in 
\cite{johnson2008}, was made, but the cool temperature of 
OGLE-2008-BLG-209S meant 
that no accurate $\teff$ could be derived this way.

\subsection{Method 4: Analysis by Cohen}

This method is also based on standard LTE assumptions, using a grid
of solar scaled ATLAS9 model stellar atmospheres from 
\cite{castelli2003}. In the analysis for OGLE-2008-BLG-209S, only 
lines with wavelengths longer than 5400\,{\AA} were selected due 
to the low $S/N$ in the crowded blue parts of the spectrum. 
Also, the stronger lines in these high gravity dwarf stars with their 
extended damping wings, which are difficult to measure in crowded 
spectra of only moderate $S/N$, are rejected. As in methods 1-3, 
damping constants from \cite{barklem4} were used when available.
Abundances were calculated with the MOOG 2002 package \citep{sneden1973}.

The effective temperature is determined by requiring excitation 
equilibrium for the set of \ion{Fe}{i} lines with measured 
equivalent widths less than 130~m{\AA}. 
The microturbulence velocity ($\xi_{\rm t}$) is solved for in the 
standard way.   
As a first pass, models with surface gravity of $\log g=4.5$, and 
with solar metallicity are used. The [Fe/H] value is then determined 
from the set of \ion{Fe}{i} lines with lower excitation potential 
greater than 4.0\,eV and equivalent widths less than 130 m{\AA} using 
that choice for $\teff$.  
The calculation is repeated with different metallicity
models if the output of the first pass differs
significantly from that of the input model atmosphere.
 The surface gravity then follows by demanding ionisation 
equilibrium between \ion{Fe}{i} and \ion{Fe}{ii}. The rationale behind 
this method is further described in Cohen et al.~(2009, in preparation),
and additional details can be found in \cite{cohen2008}.

\subsection{Comparisons of stellar parameters from methods 1--4}
\label{sec:results209}

Below we present the stellar parameters for the four stars
and compare them to the ones found by the different methods and by
\cite{johnson2007,johnson2008} and \cite{cohen2008}. All values 
are given in Table~\ref{tab:parameters}. 

\begin{table*}
\centering
\caption{
Stellar parameters as determined from the different methods. 
\label{tab:parameters}
}
\setlength{\tabcolsep}{1.5mm}
\tiny
\begin{tabular}{ccccccccccrlrcl}
\hline\hline
\noalign{\smallskip}
 Object              &
  $(V-I)_{0}$        &
  $I_{0}$            &
  $V_{0}$            & 
  $M_{\rm V}$        &  
  $d$                & 
  $T_{\rm eff}$      & 
  $\log g$           &  
  $\xi_{\rm t}$      &  
  Mass               & 
\multicolumn{2}{c}{$\log \epsilon({\rm Fe})$}       & 
 [Fe/H]              &             
  Age                \\
                     & 
 $\rm [mag]$         & 
 $\rm [mag]$         & 
 $\rm [mag]$         & 
 $\rm [mag]$         &  
 $\rm [kpc]$         & 
 $\rm [K]$           &  
 $\rm [cgs]$         &   
 $\rm [\kms]$        &    
 $\rm [M_{\odot}]$   &  
 \ion{Fe}{i}         &  
 \ion{Fe}{ii}        &
                     &  
 $\rm [Gyr]$         &
  \\  
\noalign{\smallskip}
\hline
\noalign{\smallskip}
OGLE-2008-BLG-209S &    --  &    --  & 18.35  &     3.83    &  8.0  & 5243  &  3.82  &   1.01 & 1.02 & 7.24 & 7.24 & $-0.33$ & 8.5 &  Method 1   \\
   "     &  0.73  & 17.05  & 17.68  &     3.16    &  8.0  & 5307  &  3.69  &   1.13 & 1.31 & 7.26 & 7.12 & $-0.31$ & 3.8 &  Method 2   \\
   "     &        &        &        &             &       & 5250  &  3.60  &   1.40 &      & 7.04 & 7.04 & $-0.42$ &     &  Method 3   \\
   "     &        &        &        &             &       & 5325  &  3.60  &   1.30 &      & 7.16 & 7.16 & $-0.32$ &     &  Method 4   \\
\noalign{\smallskip}
\hline
\noalign{\smallskip}
OGLE-2006-BLG-265S &    --  &    --  & 19.10  &     4.58    &  8.0  & 5486  &  4.24  &   1.17 & 1.05 & 8.02 & 8.01 & $ 0.44$ & 7.5 &  Method 1   \\
   "     &  0.68  & 18.11  & 18.69  &     4.17    &  8.0  & 5526  &  4.11  &   1.25 & 1.10 & 8.02 & 7.91 & $ 0.44$ & 7.0 &  Method 2   \\
   "     &        &        &        &     4.30    &       & 5650  &  4.40  &   1.20 &      & 8.05 & 8.07 & $ 0.56$ &     &  \cite{johnson2007}        \\
\noalign{\smallskip}
\hline
\noalign{\smallskip}
MOA-2006-BLG-099S &    --  &    --  & 19.35  &     4.83    &  8.0  & 5741  &  4.47  &   0.84 & 1.12 & 7.96 & 7.95 & $ 0.39$ &  *  &  Method 1   \\
   "     &  0.74  & 18.17  & 18.81  &     4.29    &  8.0  & 5852  &  4.32  &   1.03 & 1.18 & 7.99 & 7.81 & $ 0.42$ & 2.8 &  Method 2   \\
   "     &        &        &        &     4.50    &       & 5800  &  4.40  &   1.50 &      &      &      & $ 0.36$ &     &  \cite{johnson2008}        \\
\noalign{\smallskip}
\hline
\noalign{\smallskip}
OGLE-2007-BLG-349S &    --  &    --  & 19.35  &     4.83    &  8.0  & 5229  &  4.18  &   0.78 & 0.94 & 7.96 & 7.95 & $ 0.41$ &13.0 &  Method 1   \\
   "     &  0.78  & 18.72  & 19.40  &     4.88    &  8.0  & 5210  &  4.18  &   0.76 & 0.93 & 7.96 & 7.96 & $ 0.40$ &13.7 &  Method 2   \\
   "     &        &        &        &             &       & 5480  &  4.50  &   1.00 &      & 8.02 &      & $ 0.51$ &     &  \cite{cohen2008}   \\
\noalign{\smallskip}
\hline
\end{tabular}
\end{table*}
\begin{table*}
\centering
\caption{
Elemental abundance ratios, $[X/{\rm H}]$, as determined from the 
different methods.  
\label{tab:abundances}
}
\setlength{\tabcolsep}{0.7mm}
\tiny
\begin{tabular}{crrrrrrrrrrrrrrrrl}
\hline\hline
\noalign{\smallskip}
  Object       & 
\ion{O}{i}     &  
\ion{Na}{i}    &  
\ion{Mg}{i}    &  
\ion{Al}{i}    & 
\ion{Si}{i}    &
\ion{Ca}{i}    & 
\ion{Ti}{i}    &
\ion{Ti}{ii}   & 
\ion{Cr}{i}    &
\ion{Cr}{ii}   &
\ion{Ni}{i}    &
\ion{Fe}{i}    &
\ion{Fe}{ii}   &
\ion{Zn}{i}    &
\ion{Y}{ii}    &
\ion{Ba}{ii}   &
                \\
\noalign{\smallskip}
\hline
\noalign{\smallskip}
OGLE-2008-BLG-209S & $ 0.06$ & $-0.18$ & $ 0.01$ & $-0.04$ & $-0.13$ & $-0.15$ & $-0.14$ & $ 0.03$ & $-0.29$ & $-0.27$ & $-0.28$ & $-0.33$ & $-0.28$ & $-0.31$ & $-0.45$ & $-0.30$ & Method 1  \\
   "     & $-0.07$ & $-0.12$ & $ 0.09$ &   0.00  & $-0.15$ & $-0.08$ & $-0.09$ & $-0.05$ & $-0.23$ & $-0.37$ & $-0.27$ & $-0.31$ & $-0.40$ & $-0.37$ & $-0.54$ & $-0.32$ & Method 2  \\
   "     & $-0.08$ & $-0.28$ & $-0.14$ & $-0.09$ & $-0.25$ & $-0.30$ & $-0.26$ & $-0.04$ & $-0.33$ &   --    & $-0.36$ & $-0.42$ & $-0.44$ & $-0.36$ & $-0.77$ & $     $ & Method 3  \\
   "     & $     $ & $     $ & $     $ &         & $     $ & $     $ & $     $ & $     $ & $     $ & $     $ & $     $ & $-0.32$ & $     $ & $     $ & $     $ & $     $ & Method 4  \\
\noalign{\smallskip}
\hline
\noalign{\smallskip}
OGLE-2006-BLG-265S &   0.28  &   0.58  &   0.56  &   0.52  &   0.49  &   0.37  &   0.42  &   0.34  &   0.45  &   0.26  &   0.49  &   0.44  &   0.43  &   0.32  &   0.52  &   0.29  & Method 1  \\
   "     &   0.18  &   0.64  &   0.63  &   0.55  &   0.47  &   0.43  &   0.45  &   0.27  &   0.47  &   0.18  &   0.48  &   0.44  &   0.33  &   0.30  &   0.45  &   0.25  & Method 2  \\
   "     & $-0.03$ &   0.62  &   0.48  &   0.81  &   0.48  &   0.33  &   0.46  &         &   0.64  &         &   0.62  &   0.55  &   0.57  &         &         &   0.53  & \cite{johnson2007} \\ 
\noalign{\smallskip}
\hline
\noalign{\smallskip}
MOA-2006-BLG-099S &   0.37  &   0.46  &   0.45  &   0.42  &   0.37  &   0.31  &   0.36  &   0.45  &   0.25  &   0.28  &   0.39  &   0.39  &   0.41  &   0.47  &   0.42  &   0.37  & Method 1  \\
   "     &   0.22  &   0.56  &   0.56  &   0.48  &   0.38  &   0.42  &   0.43  &   0.36  &   0.32  &   0.16  &   0.42  &   0.42  &   0.28  &   0.44  &   0.31  &   0.36  & Method 2  \\
   "     &   0.20  &   0.45  &   0.52  &   0.47  &   0.42  &   0.25  &   0.31  &   0.14  &         &         &   0.38  &   0.36  &   0.18  &   0.08  &         & $-0.25$ & \cite{johnson2008} \\
\noalign{\smallskip}
\hline
\noalign{\smallskip}
OGLE-2007-BLG-349S & $ 0.34$ & $ 0.55$ & $ 0.43$ &   0.49  & $ 0.47$ & $ 0.31$ & $ 0.43$ & $ 0.41$ & $ 0.40$ & $ 0.41$ & $ 0.48$ & $ 0.41$ & $ 0.42$ & $ 0.49$ & $ 0.26$ & $ 0.19$ & Method 1  \\
   "     & $ 0.37$ & $ 0.54$ & $ 0.42$ &   0.48  & $ 0.48$ & $ 0.30$ & $ 0.41$ & $ 0.42$ & $ 0.39$ & $ 0.42$ & $ 0.48$ & $ 0.40$ & $ 0.44$ & $ 0.50$ & $ 0.27$ & $ 0.19$ & Method 2  \\
   "     & $ 0.14$ & $ 0.77$ & $ 0.59$ &   0.49  & $ 0.47$ & $ 0.55$ & $ 0.53$ & $ 0.67$ & $ 0.53$ & $     $ & $ 0.67$ & $ 0.51$ & $ 0.43$ & $ 0.42$ & $     $ & $ 0.38$ & \cite{cohen2008} \\
\noalign{\smallskip}
\hline
\end{tabular}
\end{table*}

\paragraph{\sl OGLE-2008-BLG-209S:}

The derived stellar parameters, $\log g=3.6$ to 3.8, and $\teff=5200$ to 
5300\,K, are consistent with OGLE-2008-BLG-209S being a subgiant star 
which also can be seen in  Fig.~\ref{fig:getage}.
Furthermore, we find that OGLE-2008-BLG-209S has a 
sub-solar metallicity of $\rm [Fe/H]\approx-0.33$ (ranging between 
$\rm [Fe/H]=-0.42$ and $\rm [Fe/H]=-0.32$ for the different methods). 
This relatively low metallicity is in stark contrast to the previous 
three Bulge dwarf stars that were all determined to have highly 
super-solar metallicities: two with $\rm [Fe/H]>0.5$ 
\citep{johnson2007,cohen2008}, and one with $\rm [Fe/H]>0.3$ 
\citep{johnson2008}. From methods 1 and 2 we find an age for 
OGLE-2008-BLG-209S ranging between 4-8\,Gyr (see Fig.~\ref{fig:getage}).

\paragraph{\sl OGLE-2006-BLG-265S:}

The effective temperature 
from the different methods and from \cite{johnson2007} agree quite 
well and range between 5500 to 5650\,K. Also the surface gravity is 
well constrained between $\log g=4.1$ to 4.4, values that are typical 
for dwarf stars. Using methods 1 and 2 we find that OGLE-2006-BLG-265S 
is a very metal-rich star at $\rm [Fe/H]=0.44$. It is however 0.12\,dex 
lower than in \cite{johnson2007} who finds $\rm [Fe/H]=0.56$. 
From methods 1 and 2 we find that OGLE-2006-BLG-265S has an age 
of approximately 7\,Gyr (see Fig.~\ref{fig:getage}).
From Fig.~\ref{fig:getage} we see that OGLE-2006-BLG-265S is a dwarf 
star close to the main sequence turn-off.

\paragraph{\sl MOA-2006-BLG-099S:}

Methods 1 and 2 give $\teff=5741$ to 5852\,K, $\log g=4.32$ to 4.47,
and $\rm [Fe/H]=0.39$ to 0.42, which are values that are very similar
to the results by \cite{johnson2008} that found $\teff=5800$\,K and 
$\rm [Fe/H]=0.36$. While \cite{johnson2008} used the microlensing
technique to estimate $M_{\rm V}$ (and thus $\log g$), they
determined the effective temperature from profile fitting of the 
wings of the H$\alpha$ and H$\beta$ Balmer lines.  Since MOA-2006-BLG-099S 
falls just outside the limits of the isochrones we can only 
give an upper limit to its age, which should be approximately 
3\,Gyr (see Fig.~\ref{fig:getage}). Given the stellar 
parameters and the position in the 
$M_{\rm V}$-$\teff$ diagram, it is clear that MOA-2006-BLG-099S is 
a main sequence dwarf star.

\paragraph{\sl OGLE-2007-BLG-349S:}

The stellar parameters based on methods 1 and 2 are very
similar: $\teff\approx5220$\,K and $\rm [Fe/H]\approx0.40$.
These results are lower than the ones found by \cite{cohen2008},
$\teff$ by $\sim 260$\,K and [Fe/H] by $\sim0.1$\,dex. Given the
uncertainties claimed by \cite{cohen2008} of 100\,K in $\teff$
and $\sim0.1$\,dex in [Fe/H], and the uncertainties
for methods 1 and 2 as listed in Sect.~\ref{sec:errors}, we see that
the metallicities are compatible and also the temperatures,
but just barely. We find that OGLE-2007-BLG-349S is a dwarf star 
close to the
main sequence turn-off, and that it has a high age of
approximately 13\,Gyr (see Fig.~\ref{fig:getage}).

\paragraph{\sl In summary,}

the differences we find
in the stellar parameters among the groups (and methods),
with the possible exception of OGLE-2007-BLG-349S,  are easily 
within the error bars (see Sect.~\ref{sec:errors}).

\subsection{Inconsistencies}
\label{sec:inconsistencies}

\paragraph{\sl Equivalent width measurements and $\log gf$-values:}

For OGLE-2008-BLG-209S, method 3 gives a $\teff$ similar to methods 
1 and 2 but a lower metallicity: 0.2\,dex in absolute values and 
0.1\,dex when normalised to the Sun. The decrease in the difference 
when normalising to the Sun can be explained as a result 
of the lower Solar abundances that method 3 produces compared to methods 
1 and 2 (see Sect.~\ref{sec:solaranalysis} and Fig.~\ref{fig:ganymede}).
The difference of 0.2\,dex in the absolute Fe abundances 
can be explained by difference of 0.2\,dex in $\log g$ and 0.4\,$\kms$ in 
$\xi_{\rm t}$. In \cite{bensby2003,bensby2005}
where we in detail investigated the effects of uncertainties in
these parameters we see that the above differences in $\xi_{\rm t}$
and $\log g$ would result in a changes of $\sim 0.10-0.15$\,dex 
and $\sim 0.03$\,dex, respectively, in the output Fe abundance. 
Therefore it is clear that it is mainly the microturbulence parameter 
that is to blame for the different absolute Fe abundances between
methods 1 and 3. In Sect.~\ref{sec:solaranalysis} we also
show that the output abundances from methods 1 and 3 are very similar
if the input equivalent widths are similar and the model stellar
parameters are the same. Hence, there are no problems with the
model stellar atmospheres (MARCS 1997 versus MARCS 2008), nor the 
different abundance programs (EQWIDTH versus TurboSpectrum), that are 
used by methods 1 and 3 in the abundance analysis.

Method 4 gives a higher $\teff$ for OGLE-2008-BLG-209S 
than methods 1, 2, and 3, and a similar metallicity.
As shown in Sect.~\ref{sec:eqws} there are significant 
differences between the \ion{Fe}{i} and \ion{Fe}{ii} equivalent 
widths as measured by Bensby and Cohen. The Bensby measurements are
4\,\% and 12\,\% larger for \ion{Fe}{i} and \ion{Fe}{ii}
lines, respectively, and can possibly affect the accuracy of 
the ionisation balance that are used in finding the stellar
parameters. To investigate if this could 
explain the differences in the stellar parameters between methods 
1 and 4 we divide the \ion{Fe}{i} and \ion{Fe}{ii} equivalent widths 
of Bensby by 1.04 and 1.12, respectively, and redo the determination
of the stellar parameters.
The effect is a slightly lower effective temperature, 5214\,K, from
method 1 and does not resolve the problem with the different
effective temperatures between methods 1 and 4. 
However, they only differ by 80\,K, which is well within the uncertainties.

In Sect.~\ref{sec:eqws} we also found differences in the $\log gf$-values 
for the \ion{Fe}{i} and \ion{Fe}{ii} lines between the Bensby and Cohen 
linelists. Average differences  are $-0.03\pm0.07$\,dex for \ion{Fe}{i} lines 
and $-0.05\pm0.09$\,dex for \ion{Fe}{ii} lines, with the values from \cite{cohen2008} 
being the larger ones. In methods 1 and 2 the abundances from \ion{Fe}{i} 
and \ion{Fe}{ii} lines have been used in their absolute form, i.e. 
they have not been normalised to the Sun, when determining the stellar 
parameters. Therefore, if the different methods/studies have different 
$\log gf$-values for the \ion{Fe}{i} and \ion{Fe}{ii} lines it may 
have an impact on the stellar parameters.  Applying similar corrections 
to the \ion{Fe}{i} and \ion{Fe}{ii} $\log gf$-values of the Bensby 
linelist, as was done above to the Bensby equivalent widths, and 
redoing the analysis have essentially no effect on the derived parameters.

Furthermore, the recent study by \cite{melendez2009} claims that there
might be problems with the \ion{Fe}{ii} $\log gf$-values 
of \cite{raassen1998} that were adopted by \cite{bensby2003} and that 
we use in methods 1 and 2. According to \cite{melendez2009} 
these $\log gf$-values are both inaccurate and imprecise. We therefore 
check our stellar parameters for OGLE-2008-BLG-209S using the 
revised \ion{Fe}{ii} $\log gf$-values by \cite{melendez2009} and we 
find slightly different values for the stellar parameters: 
$\teff=5197$\,K, $\log g=3.88$, and $\xi_{\rm t}=0.93$.
The age of the star also becomes higher, $12.5\pm6$\,Gyr compared to 
$8.5\pm6$\,Gyr. Apart from the age difference, these changes are really  
marginal. Therefore, and as about half the $\log gf$-values by
\cite{melendez2009} are based on an inverse solar analysis (which 
most likely is the cause for the claimed abundance spread decrease by
\citealt{melendez2009}),
we will keep the theoretical values by \cite{raassen1998}. 
We will then have a single source for our $\log gf$-values, and will also
be independent on the methods \cite{melendez2009} use to measure equivalent 
widths for these lines, and to their choice of model stellar
atmospheres used in the inverse solar analysis.

\paragraph{\sl Colours and temperatures:}

From the calibration by \cite{alonso1996} we check what temperature we 
should expect given the inferred colour of OGLE-2008-BLG-209S.
The Alonso calibration is however in the Johnson system only, so we 
convert the Johnson-Cousin colour of OGLE-2008-BLG-209S using the 
relation by \cite{fernie1983}, giving $(V-I)_{J}=0.924$. 
Applying Eq.~(6) in \cite{alonso1996} we then get a temperature of 
5564\,K. And what colour would be predicted for this star at T=5250?  
Using the same equations (in reverse) we get  a Johnson-Cousin
colour $(V-I)_{0}=0.83$. 
Double-checking with the colour-$\teff$ calibration of \cite{ramirez2005}
we find that 
this star should have an intrinsic Johnson-Cousin $(V-I)_{0}\approx0.82$ 
if its temperature is about 5250. Hence it seems that all is fine and the
temperatures we derive should be good.
 
However, an intrinsic colour $(V-I)_{0}\approx0.82$ means that there 
is an inconsistency in method 2, because $\teff$ is adopted in 
disagreement of this colour (0.73 compared to 0.82). This could indicate that the reddening perhaps is wrong.

But, is this discrepancy possible?  In principle yes.  
We do have an example of a source (OGLE-2008-BLG-513S) that has
0.3\,mag less extinction than the clump.  So 0.1\,mag is not
unheard of.  Nevertheless, this discrepancy is still bigger
than what has been seen for the other three dwarf stars.    
Anyway, given the estimated uncertainties in $\teff$ (see 
Sect.~\ref{sec:errors}) it seems that an effective 
temperature of 5243\,K is not unreasonable for OGLE-2008-BLG-209S.

\section{Solar analysis}
\label{sec:solaranalysis}

\begin{figure}
\resizebox{\hsize}{!}{
\includegraphics[angle=-90,bb=28 28 570 600,clip]{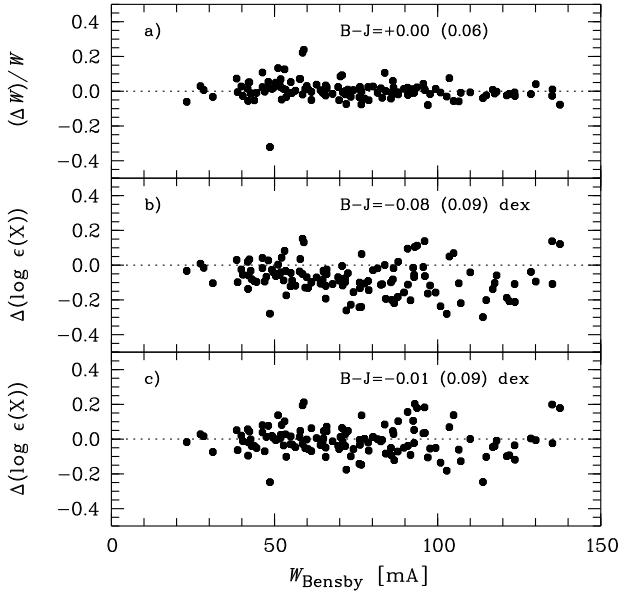}}
\caption{Differences in measured equivalent widths and elemental
abundances for the Sun for the Bensby (B) and Johnson (J) analyses of the
Ganymede solar spectrum. a) shows the
On average the Bensby equivalent widths are
$0.4\pm10$\,\% larger than Johnson's (see plot a), and the corresponding
Bensby abundances are on average $0.077\pm0.095$\,dex lower than Johnson's
(see plot b). Plot c) then shows the differences when the Bensby abundances
are computed using exactly the same stellar parameters as used
by Johnson (see discussion in text). The differences then decrease to 
less than 0.01\,dex.
Differences and standard deviations (in parentheses) are indicated
in the figures and are based on 128 spectral lines that Bensby
and Johnson have in common.
        \label{fig:ganymede}}
\end{figure}

For methods 1 and 2 a solar analysis was performed on a spectrum of 
Jupiter's moon  Ganymede that was obtained in March 2007 with the MIKE 
spectrograph using the same instrument settings as for the 
observations of OGLE-2008-BLG-209S. Analysing the solar spectrum in 
a similar way (as we know the absolute magnitude of the Sun 
we do not require ionisation equilibrium) 
as for methods 1 and 2, we derive $\teff=5790$\,K, 
$\log g=4.45$, $\xi_{\rm t}=1.04\,\kms$, and $\log\,\epsilon({\rm Fe})=7.56$ 
for the Sun. The final 
elemental abundances based on methods 1 and 2, will be normalised 
to those of the Sun from this analysis. The normalisation will be
done on a line-by-line basis, and then averaged for 
each element, making the results strictly differential to the Sun. 
This way of normalising the abundances neutralises uncertainties and 
errors in the $\log gf$-values.
It should be noted that when a line in the Solar spectrum could not be 
measured, or, in the case of Fe lines, the line strength exceeded
90\,m{\AA}, the average abundance from all the other lines of the
same species were used for the solar abundance of that line. These
cases are marked by "flag=1" in Table~\ref{tab:onlinebensby}.

In Table~\ref{tab:solarabund} we give the average solar abundances
for each species as derived by Bensby from the MIKE Ganymede spectrum. 
For comparison purposes we also give the standard
Solar abundances as given by \cite{asplundgrevessesauval2005} and
\cite{grevesse2007}. Generally our Solar abundances agree within 
0.1\,dex of the standard ones, with O and Ba being the exceptions.
The differences, all though they are small, illustrates the importance of 
doing a strictly differential analysis toward the Sun, deriving 
your own Solar abundances instead of adopting the tabulated ones.

For method 3, a solar analysis was performed using the same 
spectrum of Ganymede as was used for methods 1 and 2. 
The line list and line parameters were the same, and a
model stellar atmosphere for the Sun was 
interpolated in the same grid of model stellar atmospheres
as was used for OGLE-2008-BLG-209S, using $\teff=5770$, 
$\log g=4.44$, $\rm [Fe/H]=0$ 
($\log\,\epsilon({\rm Fe})=7.45$), and $\xi_{\rm t}=1.1\,\kms$.

For method 4, the same set of lines as used for OGLE-2008-BLG-209S
were analysed in the Solar spectrum\footnote{Available at
{\tt ftp://nsokp.nso.edu/pub/atlas/visatl}.} 
by \cite{wallace1998}. These were then use to determine [X/H] for 
each species for OGLE-2008-BLG-209S.

As the same Ganymede spectrum was used by both Bensby and Johnson to 
determine the Solar elemental abundances we show in 
Figure~\ref{fig:ganymede}a a comparison of measured equivalent 
widths and absolute elemental abundances for 128 spectral lines
in common. The equivalent width measurements are in good 
agreement and differs by only 0.5\%, with the measurements
by Bensby being the slightly larger ones.  Surprisingly, this
good agreement in the equivalent widths is not reflected in the 
derived elemental abundances for the Sun. On average the abundances 
from Bensby's analysis are 0.08\,dex {\it lower} than the
abundances from Johnson's analysis (see Fig.~\ref{fig:ganymede}b). 
This is puzzling, but it turns out that the solar models
used by Bensby and Johnson differ. While Johnson used  a MARCS model
with $\log\,\epsilon({\rm Fe})=7.45$,  Bensby used a MARCS model
with $\log\,\epsilon({\rm Fe})=7.56$, and of course the values for the
microturbulence parameter also differed slightly. So, redoing the
Bensby solar abundances, using exactly the same set of model
stellar parameters as used by Johnson, the differences in the solar 
abundances become less than 0.01\,dex  (see Fig.~\ref{fig:ganymede}c).
This also explains why the differences between methods 1 and 3 in the 
Fe abundances are different depending on if comparisons are made 
between absolute abundances or normalised abundances 
(see Table~\ref{tab:parameters}). As we are primarily interested
in comparing stars with each other it is []-notation that is relevant 
here and it is encouraging to see how well the different methods do 
reproduce the results (once any differences in the methodology are 
fully accounted for). This is perhaps not surprising but worth noting 
and serves as a reminder that if one attempts to combine data from 
different studies then careful normalisation is a must. 

Equivalent widths and absolute elemental abundances (also for the Sun) 
and normalised abundances are given for individual lines
are given in Table~\ref{tab:onlinebensby}.
Average elemental abundance ratios are given in Table~\ref{tab:abundances}.
Absolute and normalised abundances from the \ion{Fe}{i}
and \ion{Fe}{ii} lines are also given in Table~\ref{tab:parameters}.

\begin{table}
\centering
\caption{
Our solar abundances based on observations
of Jupiter's moon Ganymede. Last column give
the standard Solar photospheric abundances
given by \cite{asplundgrevessesauval2005}. 
\label{tab:solarabund}
}
\tiny
\begin{tabular}{lcrc}
\hline\hline
\noalign{\smallskip}
Ion             &
$\log \epsilon (X)_{\odot}$ &
$N$             &
A2005           \\
\noalign{\smallskip}
\hline
\noalign{\smallskip}
\ion{Fe}{i}  & $7.57\pm0.08$ & 148  & 7.45  \\
\ion{Fe}{ii} & $7.52\pm0.13$ &  17  & 7.45  \\
\ion{O}{i}   & $8.86\pm0.02$ &   3  & 8.66  \\
\ion{Na}{i}  & $6.29\pm0.08$ &   4  & 6.17  \\
\ion{Mg}{i}  & $7.62\pm0.08$ &   5  & 7.53  \\
\ion{Al}{i}  & $6.48\pm0.07$ &   7  & 6.37  \\
\ion{Si}{i}  & $7.60\pm0.07$ &  27  & 7.51  \\
\ion{Ca}{i}  & $6.32\pm0.11$ &  19  & 6.31  \\
\ion{Ti}{i}  & $4.93\pm0.11$ &  22  & 4.90  \\
\ion{Ti}{ii} & $4.90\pm0.17$ &  14  & 4.90  \\
\ion{Cr}{i}  & $5.55\pm0.06$ &   6  & 5.64  \\
\ion{Cr}{ii} & $5.69\pm0.04$ &   5  & 5.64  \\
\ion{Ni}{i}  & $6.22\pm0.07$ &  42  & 6.23  \\
\ion{Zn}{i}  & $4.60\pm0.08$ &   3  & 4.60  \\
\ion{Y}{ii}  & $2.12\pm0.11$ &   4  & 2.21  \\
\ion{Ba}{ii} & $2.42\pm0.07$ &   4  & 2.17  \\
\noalign{\smallskip}
\hline
\end{tabular}
\end{table}

\section{Errors}
\label{sec:errors}

\subsection{Random errors in stellar parameters}

We are performing standard LTE abundance analyses and the 
associated errors and uncertainties for the stellar parameters
should be of standard nature. Methods 1 and 2 are similar to the 
analysis carried out in \cite{bensby2003,bensby2005} where errors of 
80\,K in $\teff$ and 0.1\,dex in $\log g$ are quoted. The line-to-line
scatter in those studies for [Fe/H] are of the order 0.08\,dex.
For OGLE-2008-BLG-209S the standard deviation of the mean abundance in the \ion{Fe}{i}
abundances are about double that, $\rm \sigma [Fe/H]=0.14$ 
(see Table~\ref{tab:finalabundances}), which most likely is an 
effect of the lower quality of the spectrum as compared to the spectra
used in \cite{bensby2003,bensby2005} that had $S/N\gtrsim 200$.
Therefore we estimate that the uncertainties in $\teff$ and $\log g$
should be higher for  OGLE-2008-BLG-209S. A conservative estimate
is 0.2\,dex in $\log g$ and 200\,K in $\teff$. The same numbers should
hold for OGLE-2006-BLG-265S and MOA-2006-BLG-099S that have
errors in the \ion{Fe}{i} abundances comparable
to OGLE-2008-BLG-209S (see Table~\ref{tab:finalabundances}). 
The spectrum of OGLE-2007-BLG-349S is however of significantly 
higher quality than the others, which also is reflected in the lower 
\ion{Fe}{i} standard deviation around the mean abundance, 0.10\,dex 
(see Table~\ref{tab:finalabundances}). Hence, for OGLE-2007-BLG-349S 
we estimate that the  errors in $\teff$, $\log g$ are of the order 
150\,K and 0.15\,dex, respectively.
The magnitudes of these errors are in line with the values given by
\cite{johnson2007,johnson2008} and \cite{cohen2008}
for OGLE-2006-BLG-265S, MOA-2006-BLG-099S, and OGLE-2007-BLG-349S.

In the last column of Table~\ref{tab:finalabundances} we list by 
how much the [$X$/Fe] ratios are affected by these
uncertainties in the stellar parameters. The values are taken
as double the values as listed in \cite{bensby2005} and 
\cite{bensby2004} (for [O/Fe]) where we investigated this in detail
(see discussion above).

\subsection{Elemental abundances and the robustness of abundance ratios}
\label{sec:robustness}

We find that the 
$\rm [X/Fe]$ ratios generally agree well between the different 
methods/studies, and it is essentially only oxygen, zinc, yttrium, 
and barium that show 
any real discrepancies for one or two stars. The big difference in
$\rm [O/Fe]$ that can be seen compared to \cite{johnson2007} for
OGLE-2006-BLG-265S is clearly an effect of the almost 200\,K 
difference in effective temperature. 

\begin{figure}
\resizebox{\hsize}{!}{
\includegraphics[angle=-90,bb=65 45 550 500,clip]{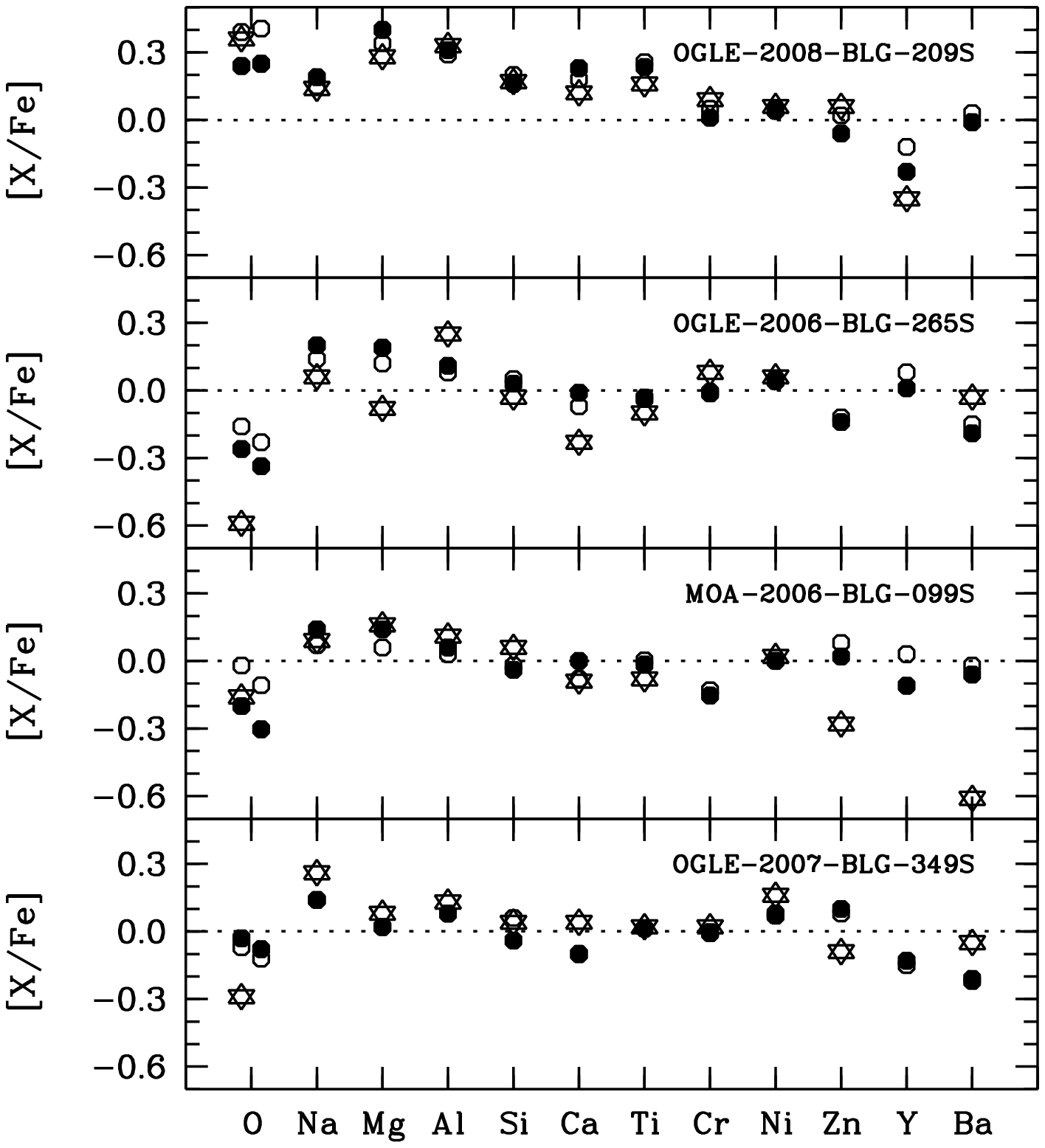}}
\caption{Comparisons of derived $\rm [X/Fe]$ ratios for method 1
(open circles), method 2 (filled circles), and from
\cite{johnson2007,johnson2008} and \cite{cohen2008} (stars). F
or OGLE-2008-BLG-209S the stars
show the values from method 3. For oxygen two values
are given side by side: on the left hand side the LTE value as it 
comes of the analysis, and on the right hand side the NLTE value 
corrected by the empirical NLTE correction formula from \cite{bensby2004}.     
\label{fig:xfe_check}}
\end{figure}

The Ba abundances from
\cite{johnson2007,johnson2008} are based on line synthesis,
and hence the equivalent widths can not be directly compared.
However, except for MOA-2006-BLG-099S, where the \cite{johnson2008} [Ba/Fe] 
ratio is extremely low, the agreement is good. The low value of 
MOA-2006-BLG-099S is also not recognised in OGLE-2006-BLG-265S, 
a similarly metal-rich star. As mentioned in \cite{johnson2008}
the reason for the low [Ba/Fe] value of MOA-2006-BLG-099S was the high
value they measured in the Sun. 
The good agreement for the other stars between Ba as derived by 
line synthesis and through equivalent width measurement
analysis illustrates the point raised by \cite{bensby2005}
that because the dominant isotopes for Ba are the even isotopes,
which do not have hyperfine structure, the inclusion of HFS
does not substantially change the abundances.
Furthermore, the line synthesis of the \ion{Ba}{ii} line at 6141\,{\AA}
done by Johnson in method 3 for OGLE-2008-BLG-209S
included an \ion{Fe}{i} line that blends with the feature. However, this
line contributes very little to the line profile. A
synthesis of the line showed that the equivalent width 
increased from 127.9\,nm to 132.7\,nm if the
\ion{Fe}{i} line was included. Therefore, using the equivalent width 
and assuming that it is all Ba leads to a very small change in the
Ba abundance.

\section{Bulge membership for OGLE-2008-BLG-209S}
\label{sec:bulgemember}

Bulge microlensing sources tend to be slightly biased to be more 
distant than the mean distance to the Bulge (``the Bulge clump"),
see e.g. \cite{kane2006}.  
The lens must obviously be in front of the source, so if the lens is 
in the Bulge (which most are), the source must be drawn from the Bulge 
stars behind it, so on average more distant than the Bulge centre.  
In fact, the bias is stronger than that because the probability of 
lensing goes basically as $\sqrt{(d_{source} - d_{lens})}$ for 
lenses and source in the Bulge. For disc lenses (the minority) there 
is no significant bias. This bias would tend to make the source more 
luminous than the microlens-model based estimate based on assuming it 
is at the same distance as the red clump.  This goes in  
the opposite direction of the difference between the spectroscopic
and microlens determinations of $M_{\rm V}$.  

Based on our spectroscopic parameters from method 3, $\teff=5250$ 
and $\log g=3.6$, we use the Y$^{2}$-isochrones at $Z=0.007$ to 
estimate an $M_{\rm I}$ of 2.5. Combined with the $I_{0}=17.05$ 
microlensing estimate, we derive a distance to  OGLE-2008-BLG-209S 
of 8\,kpc. Now, if we take $M_{\rm V}=3.83$ from Table~\ref{tab:parameters},
(method 1) and $V_{\rm 0}=17.68$ (microlensing estimate from method 2), 
the distance comes out to 6\,kpc. The difference is probably due to 
the higher gravity (i.e., fainter absolute magnitude). Since the 
errors on the gravity are not insignificant (between the small 
number of \ion{Fe}{ii} lines and the correlation between temperature 
and gravity, not to mention systematic uncertainties between isochrone 
and spectroscopic determinations), there are large errors attached to 
both these distances.

However, the high heliocentric radial velocity of OGLE-2008-BLG-209S 
(see Sect.~\ref{sec:rv}) demonstrates that it is likely to be in the 
Bulge, and the relatively high radial velocities of the other three 
dwarf stars also strongly suggest Bulge membership. 
This high velocity dispersion of our four stars
is also in agreement with what is seen in large samples
of giant stars in the Bulge \citep[e.g.][]{sadler1996,zoccali2008}. 

In conclusion, based on the radial velocity, probability
of microlensing, and small distance from the Galactic plane, 
we expect this star to be in the Bulge. This is consistent with the 
range of distances derived from spectroscopic parameters and isochrones.

\begin{table}
\centering
\caption{
Final elemental abundance ratios, $[X/{\rm Fe}]$,
based on method 1. For each abundance ratio we 
also give the standard deviation of the mean abundance 
(added in quadrature for the two elements used
to construct the abundance ratio), and the number
of lines used. These are the values used in subsequent
abundance plots. The last column gives an estimate how the
abundance ratios are affected by random errors in the stellar
parameters. The adopted uncertainties for the Bulge dwarf
stars are two times the values as given in \cite{bensby2005} 
and \cite{bensby2004}.
\label{tab:finalabundances}
}
\setlength{\tabcolsep}{1.7mm}
\tiny
\begin{tabular}{lrrrrc}
\hline\hline
\noalign{\smallskip}
        & BLG-209S & BLG-265S & BLG-099S & BLG-349S  & $\langle \sigma_{\rm rand} \rangle$ \\
\noalign{\smallskip}
\hline
\noalign{\smallskip}
[Fe/H]   & $-0.33$  &   0.44   &   0.39   &   0.41  & 0.12 \\
         &   0.14   &   0.12   &   0.15   &   0.10  &      \\
         &    146   &     92   &    119   &    103  &      \\
\noalign{\smallskip}
[O/Fe]   &   0.39   & $-0.16$  & $-0.02$  & $-0.07$ & 0.24 \\
         &   0.17   &   0.13   &   0.15   &   0.12  &      \\
         &      3   &      3   &      3   &      3  &      \\
\noalign{\smallskip}
[Na/Fe]  &   0.15   &   0.14   &   0.07   &   0.14  & 0.06 \\
         &   0.16   &   0.14   &   0.17   &   0.16  &      \\
         &      4   &      4   &      4   &      4  &      \\
\noalign{\smallskip}
[Mg/Fe]  &   0.34   &   0.12   &   0.06   &   0.02  & 0.12 \\
         &   0.14   &   0.17   &   0.17   &   0.12  &      \\
         &      5   &      2   &      4   &      3  &      \\
\noalign{\smallskip}
[Al/Fe]  &   0.29   &   0.08   &   0.03   &   0.08  & 0.10 \\
         &   0.17   &   0.18   &   0.17   &   0.15  &      \\
         &      7   &      3   &      4   &      3  &      \\
\noalign{\smallskip}
[Si/Fe]  &   0.20   &   0.05   & $-0.02$  &   0.06  & 0.10 \\
         &   0.18   &   0.19   &   0.20   &   0.14  &      \\
         &     27   &     23   &     25   &     23  &      \\
\noalign{\smallskip}
[Ca/Fe]  &   0.18   & $-0.07$  & $-0.08$  & $-0.10$ & 0.06 \\
         &   0.20   &   0.17   &   0.19   &   0.14  &      \\
         &     19   &     16   &     18   &     18  &      \\
\noalign{\smallskip}
[Ti/Fe]  &   0.26   & $-0.04$  &   0.00   &   0.01  & 0.10 \\
         &   0.22   &   0.15   &   0.23   &   0.17  &      \\ 
         &     36   &     13   &     30   &     24  &      \\
\noalign{\smallskip}
[Cr/Fe]  &   0.05   & $-0.01$  & $-0.13$  &   0.00  & 0.10 \\
         &   0.20   &   0.13   &   0.19   &   0.12  &      \\
         &     10   &      8   &     12   &      7  &      \\
\noalign{\smallskip}
[Ni/Fe]  &   0.05   &   0.05   &   0.00   &   0.07  & 0.04 \\
         &   0.23   &   0.18   &   0.22   &   0.14  &      \\
         &     40   &     30   &     37   &     39  &      \\
\noalign{\smallskip}
[Zn/Fe]  &   0.02   & $-0.12$  &   0.08   &   0.08  & 0.12 \\
         &   0.34   &   0.58   &   0.20   &   0.18  &      \\
         &      3   &      2   &      3   &      3  &      \\
\noalign{\smallskip}
[Y/Fe]   & $-0.12$  &   0.08   &   0.03   & $-0.15$ & 0.16 \\
         &   0.18   &   0.14   &   0.38   &   0.16  &      \\
         &      4   &      2   &      5   &      5  &      \\
\noalign{\smallskip}
[Ba/Fe]  &   0.03   & $-0.15$  & $-0.02$  & $-0.22$ & 0.12 \\
         &   0.16   &   0.17   &   0.17   &   0.17  &      \\
         &      4   &      3   &      4   &      4  &      \\
\noalign{\smallskip}
\hline
\end{tabular}
\end{table}

\section{Stellar age and stellar mass}

\begin{figure*}
\resizebox{\hsize}{!}{
\includegraphics[angle=-90,bb=240 28 570 435,clip]{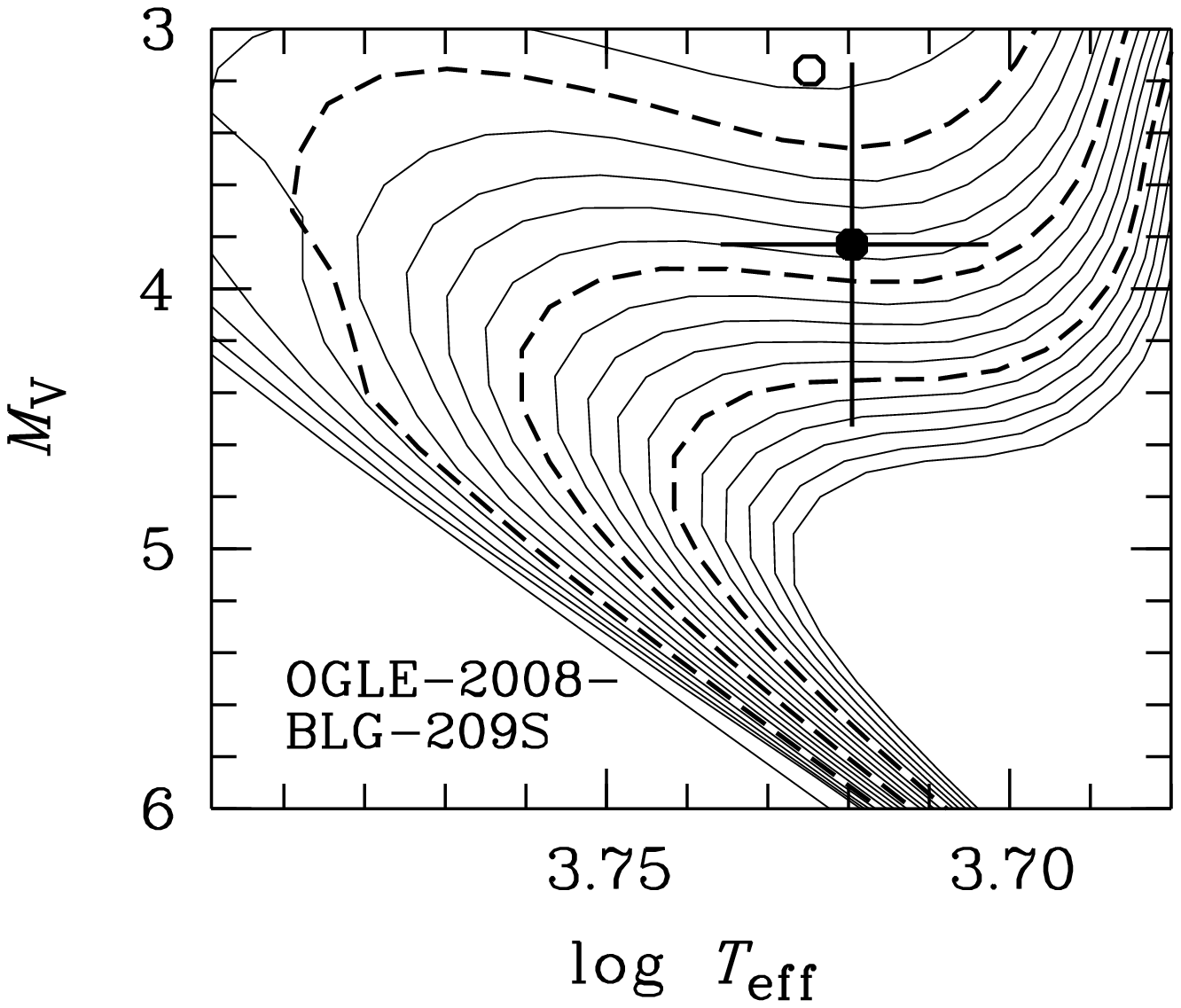}
\includegraphics[angle=-90,bb=240 110 570 435,clip]{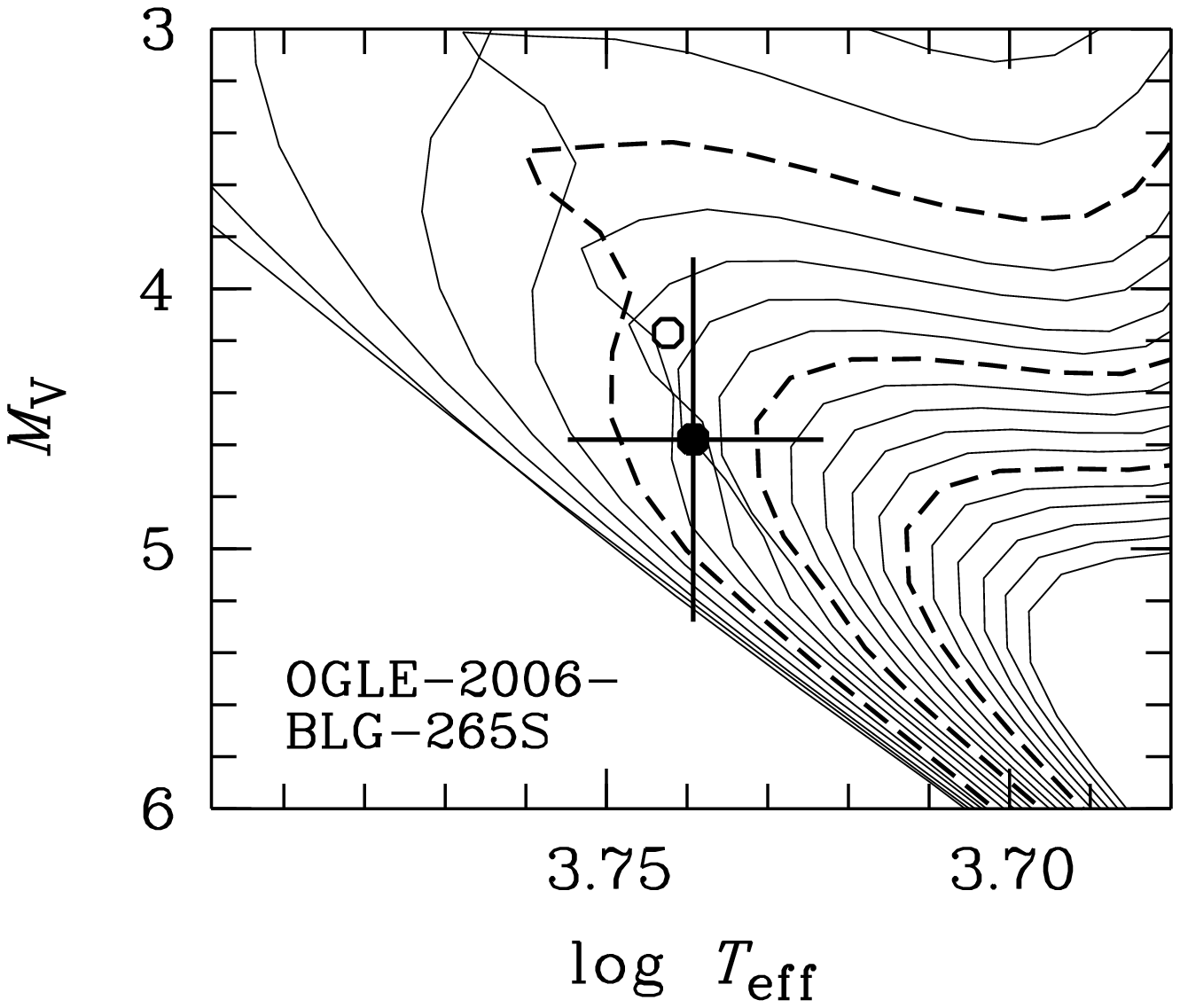}
\includegraphics[angle=-90,bb=240 110 570 435,clip]{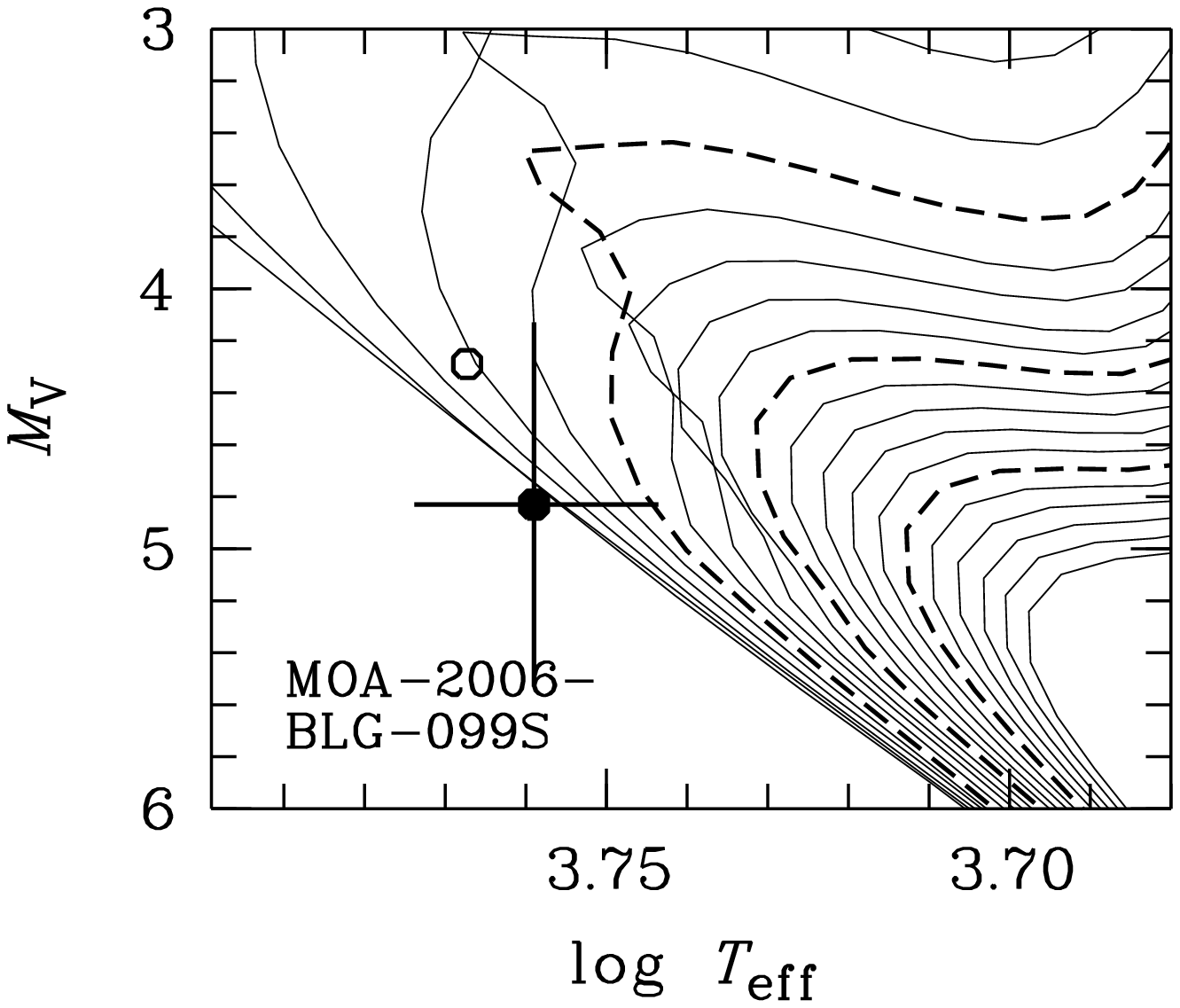}
\includegraphics[angle=-90,bb=240 110 570 450,clip]{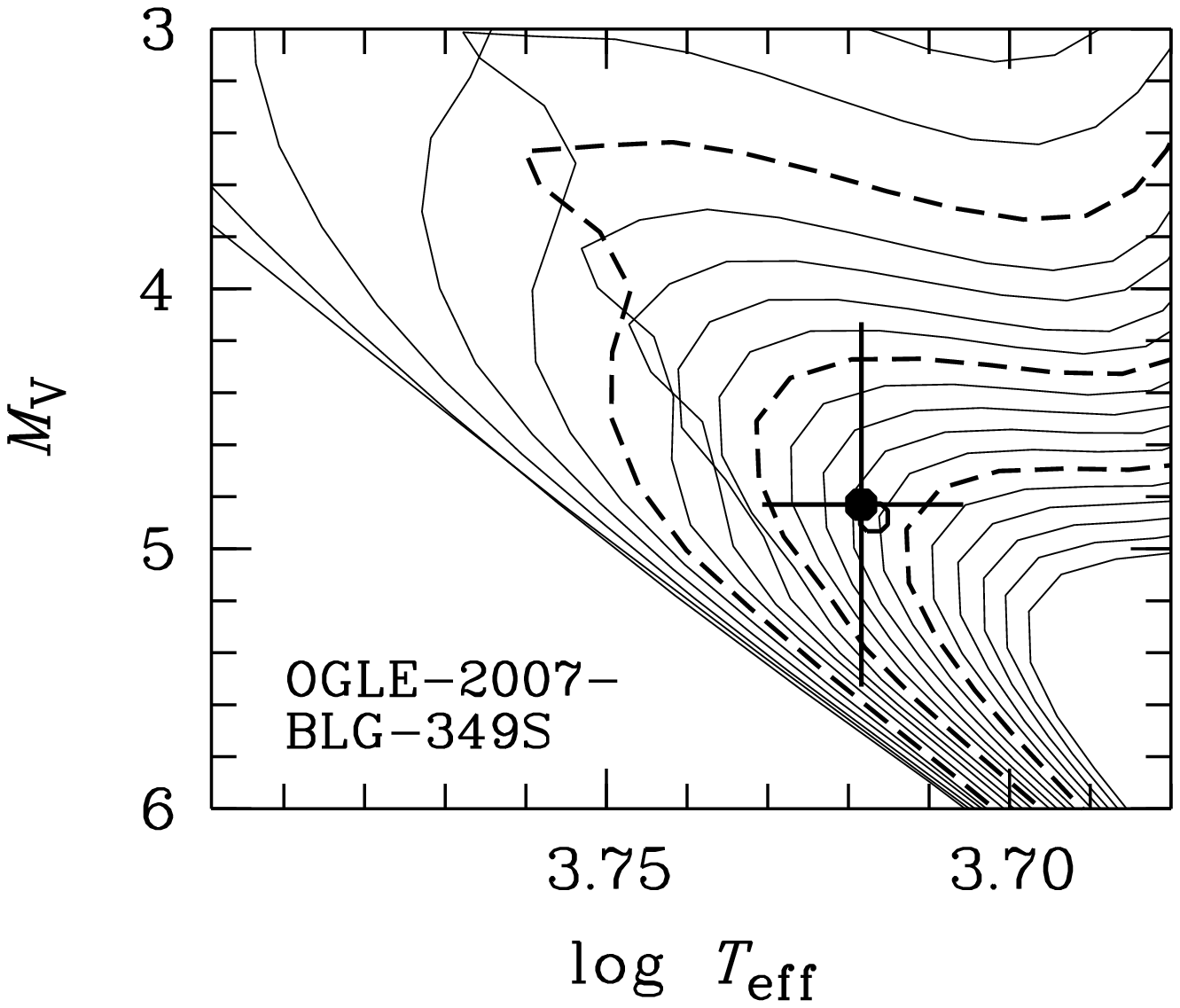}}
\caption{
         The Yonsei-Yale (Y$^{2}$) isochrones \citep{yi2001,kim2002,demarque2004} 
         that were used to estimate the stellar ages. Isochrones are plotted 
         in 1\,Gyr steps, with every 5\,Gyr isochrone in dashed lines. For each 
         target we used isochrones metallicities according to 
         Table~\ref{tab:parameters}, and for OGLE-2008-BLG-209S 
         also with an appropriate the $\alpha$-enhancement. 
         Filled circles
         represent the stellar parameters derived from method 1
         and open circles the stellar parameters from 
         method 2. 
         Error bars indicate uncertainties of 0.7\,mag in $M_{\rm V}$
         and 200\,K in $\teff$.        
         \label{fig:getage}}
\end{figure*}

When the final values of $\log g$, $\teff$, and $\xi_{\rm t}$ have been 
found we use the fundamental astrophysical relation 
$g \propto \mathcal{M}\cdot \teff^{4}/L$  to 
estimate the luminosity (and hence the apparent magnitude),
assuming a distance of 8\,kpc to the star.
For methods 1 and 2 stellar ages were then estimated using the Yonsei-Yale 
isochrones \citep{yi2001,kim2002,demarque2004}. Individual sets of 
isochrones with appropriate metallicities and $\alpha$-enhancements 
(in the case of OGLE-2008-BLG-209S) were calculated for the different 
stars. Ages were then read off from the best fitting isochrone
in the $\log\teff$-$M_{\rm V}$ plane (see
Fig.~\ref{fig:getage}). Upper and lower limits to the ages were 
estimated from the error bars based on the uncertainties in $\teff$ 
and $M_{\rm V}$. For $\teff$ we adopt error bars of 200\,K for 
OGLE-2008-BLG-209S, OGLE-2006-BLG-265S, and MOA-2006-BLG-099S, and 
150\,K for OGLE-2007-BLG-349S (see Sect.~\ref{sec:errors}). The 
errors in $M_{\rm V}$ are difficult to estimate. Given the range of values 
by the different methods in Table~\ref{tab:parameters} we adopt a 
value of 0.7\,mag for the error in $M_{\rm V}$ for all four 
stars. This value is consistent with an uncertainty of 2\,kpc in the adopted 
distance (see Sect.~\ref{sec:bulgemember}). 
These error bars are shown in the isochrone plots (see 
Fig.~\ref{fig:getage}). 

Also, stellar masses ($\mathcal{M}$) were determined from the evolutionary 
tracks by \cite{yi2003} and are given in Table~\ref{tab:parameters}.

\section{Stellar populations}
\label{sec:discussion}

Even if there are some discrepancies in the stellar
parameters from the different methods that
we have presented, they do produce results that are very similar 
in terms of abundance ratios. This is an encouraging result as it 
is the
abundance ratios that are the main results used in discussions of
the enrichment of heavy elements in a stellar population. 

However, in order to put the new results for these four Bulge stars 
into context we will adopt the results from method 1.
Method 1 is the method that most closely resembles
the method used in
\cite{bensby2003,bensby2005} and Bensby et al.~(2009, in prep),
and will enable a differential comparison of the Bulge stars
with a large sample of dwarf and subgiant stars in the nearby thin and 
thick discs. This choice is for
consistency, and importantly,
it does not make any judgement regarding which
set of abundances is the most accurate.

\begin{figure*}
\resizebox{\hsize}{!}{\includegraphics[angle=-90,bb=360  40 510 430,clip]{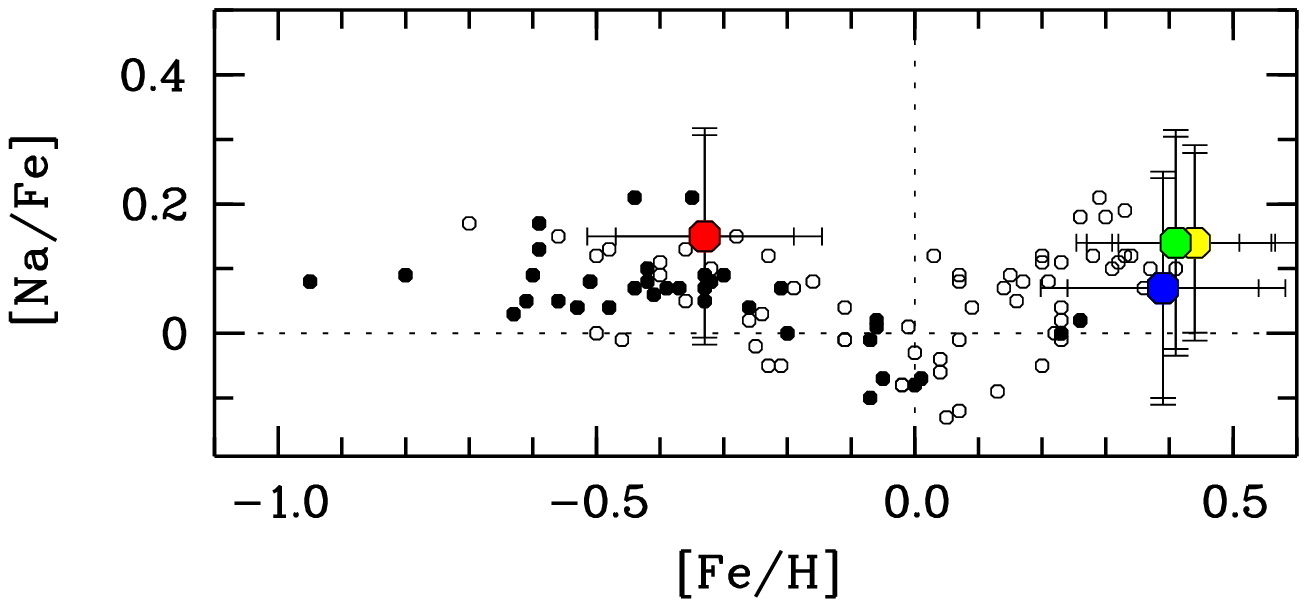}
                      \includegraphics[angle=-90,bb=360  40 510 440,clip]{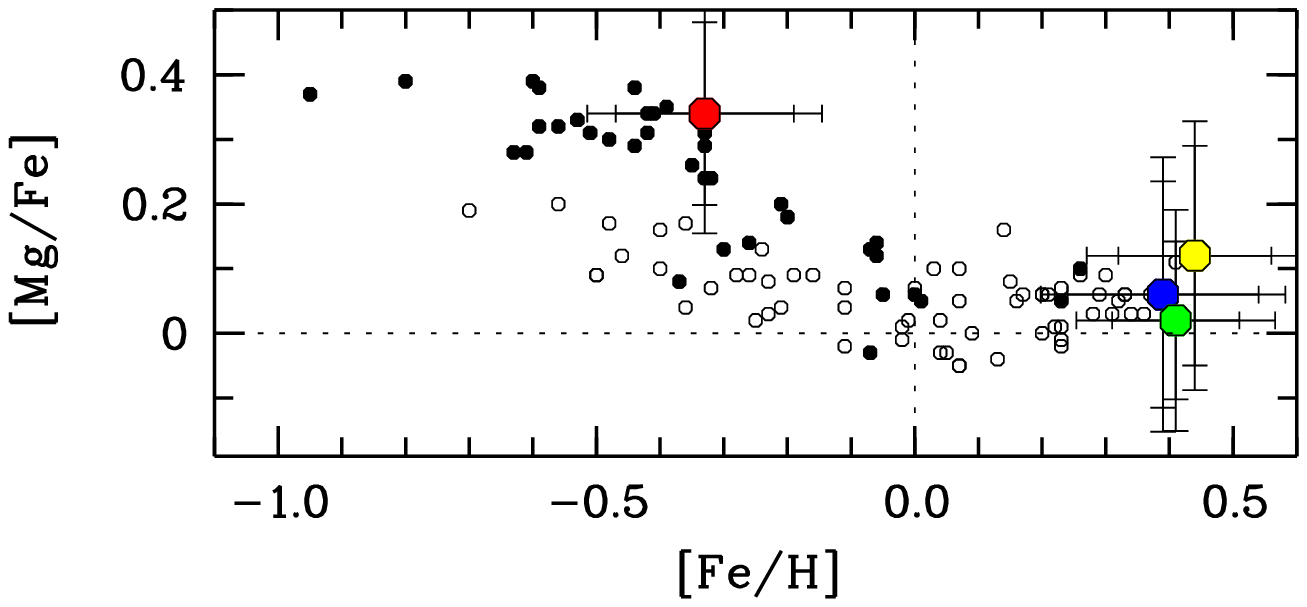}}
\resizebox{\hsize}{!}{\includegraphics[angle=-90,bb=373  40 510 430,clip]{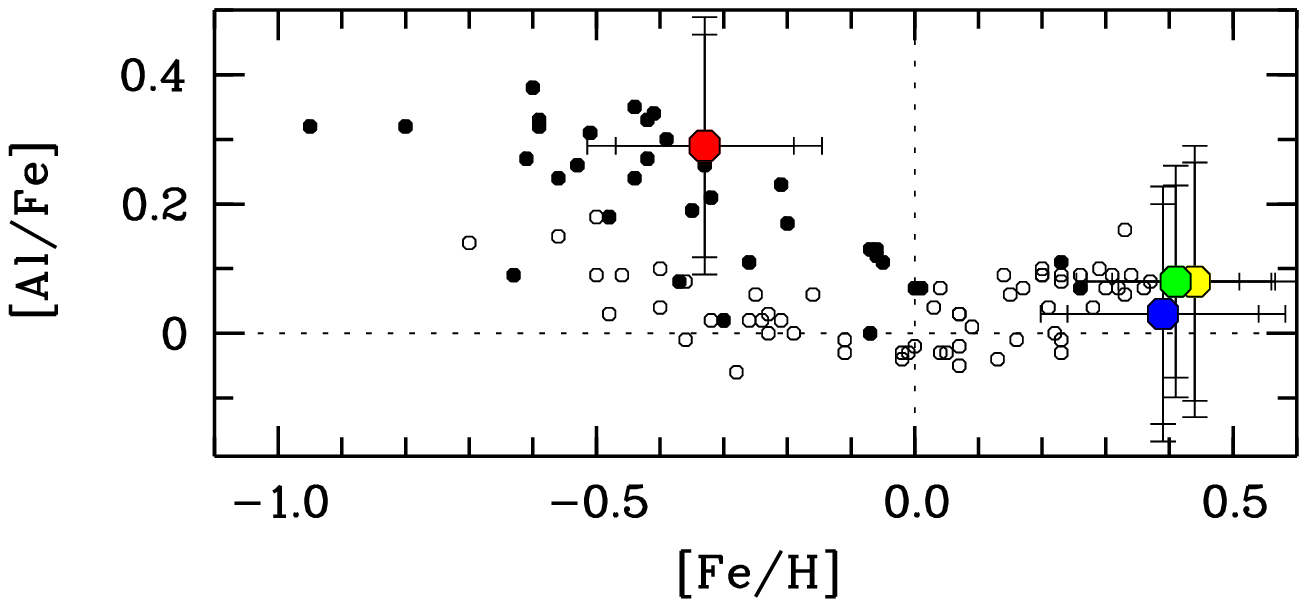}
                      \includegraphics[angle=-90,bb=373  40 510 440,clip]{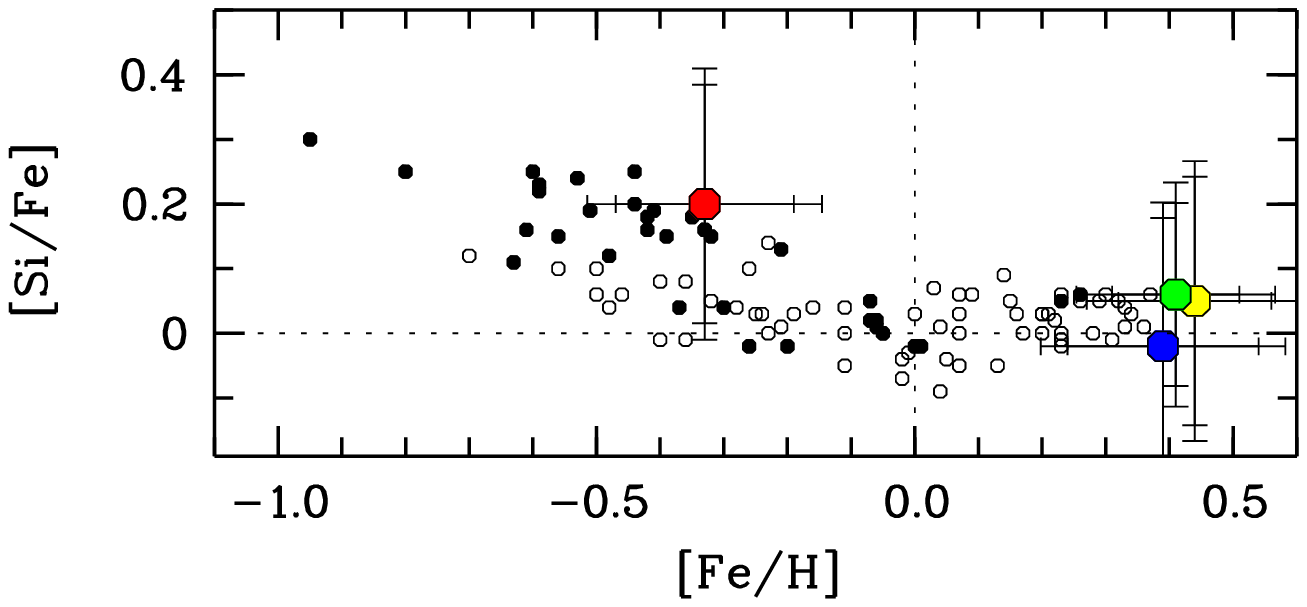}}
\resizebox{\hsize}{!}{\includegraphics[angle=-90,bb=373  40 510 430,clip]{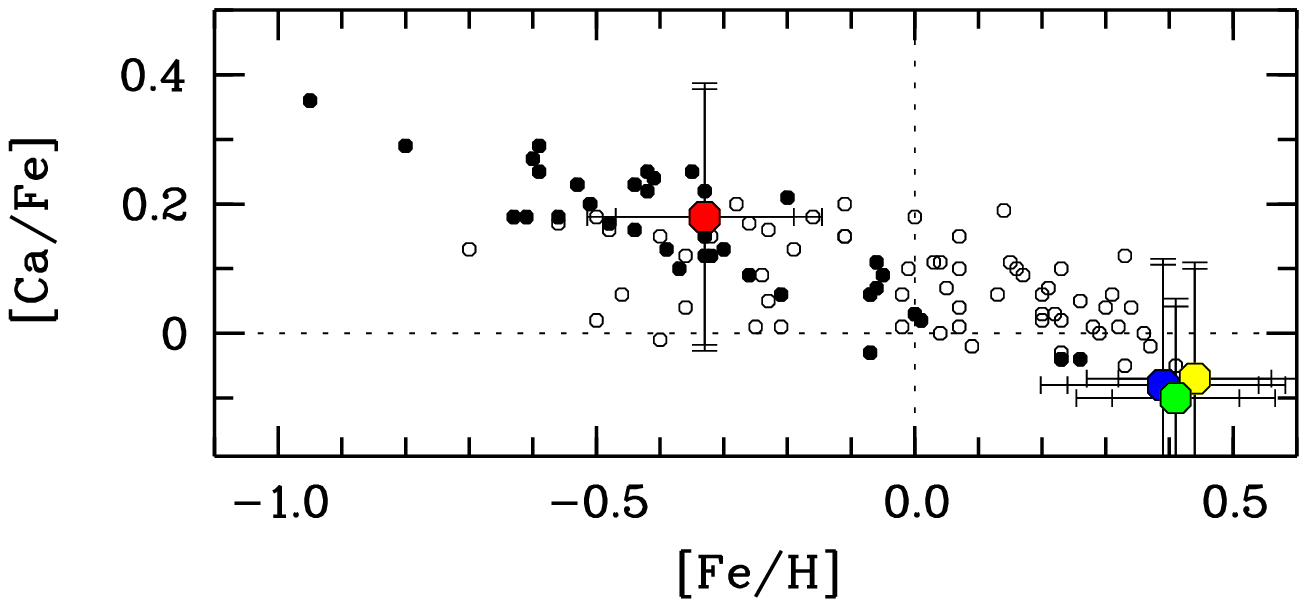}
                      \includegraphics[angle=-90,bb=373  40 510 440,clip]{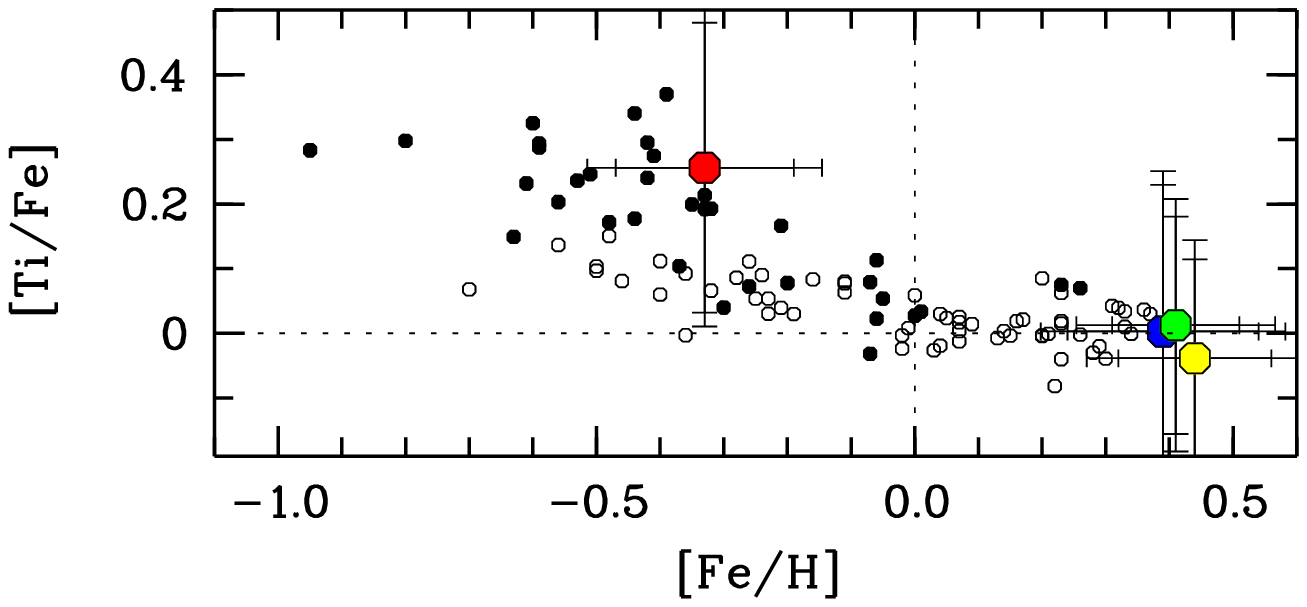}}
\resizebox{\hsize}{!}{\includegraphics[angle=-90,bb=390  40 510 430,clip]{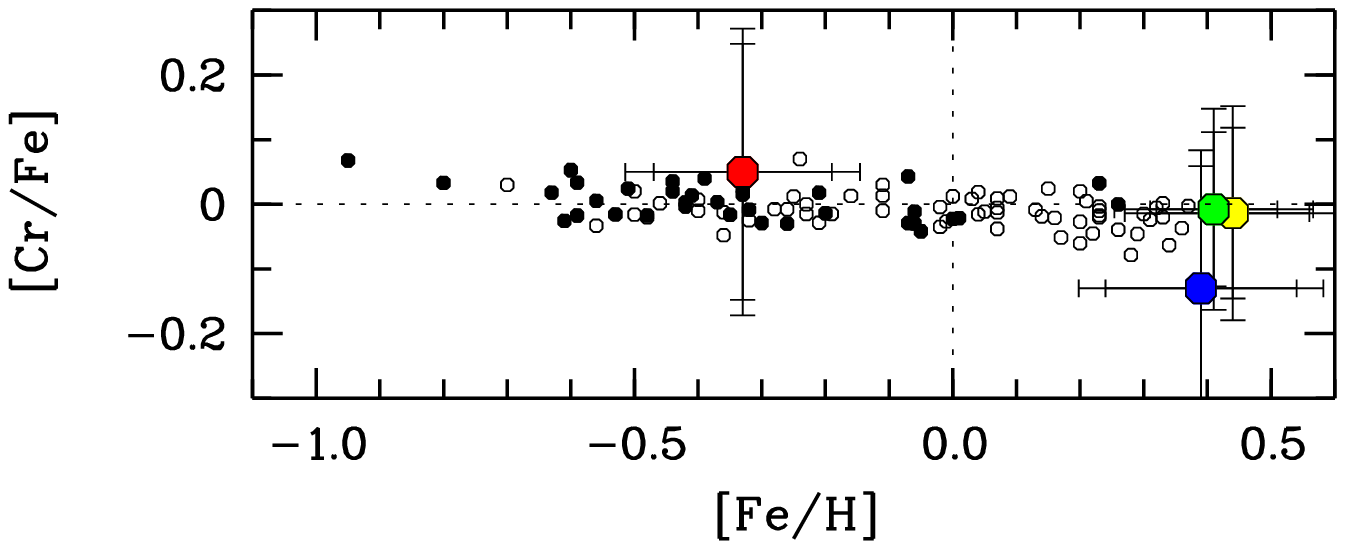}
                     \includegraphics[angle=-90,bb=390  40 510 440,clip]{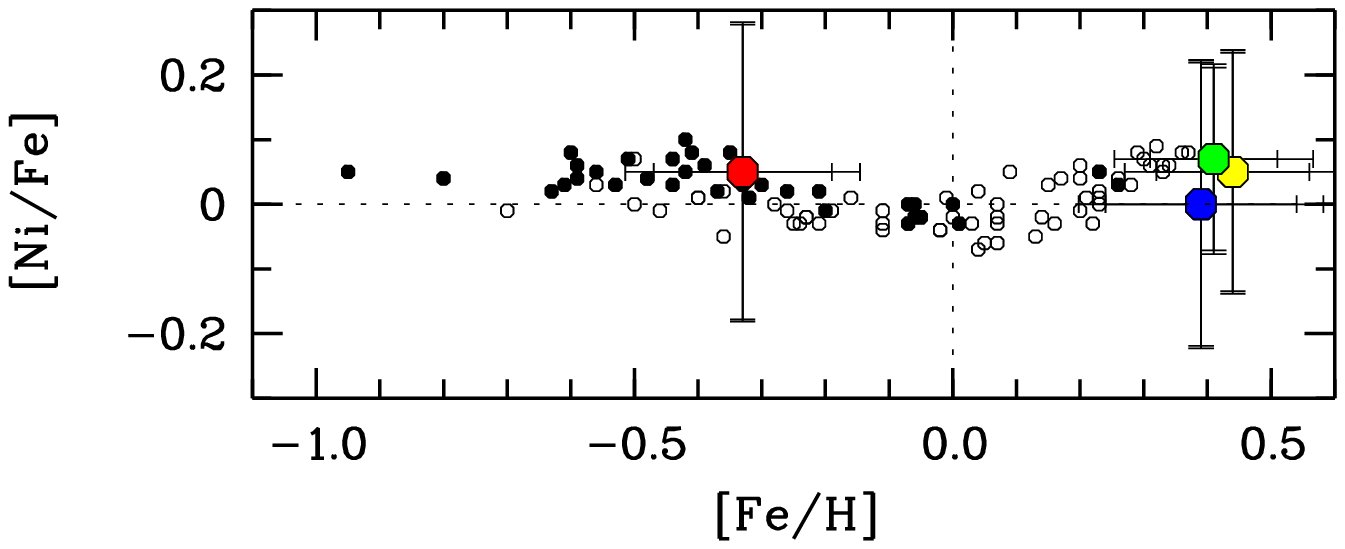}}
\resizebox{\hsize}{!}{\includegraphics[angle=-90,bb=315  40 510 430,clip]{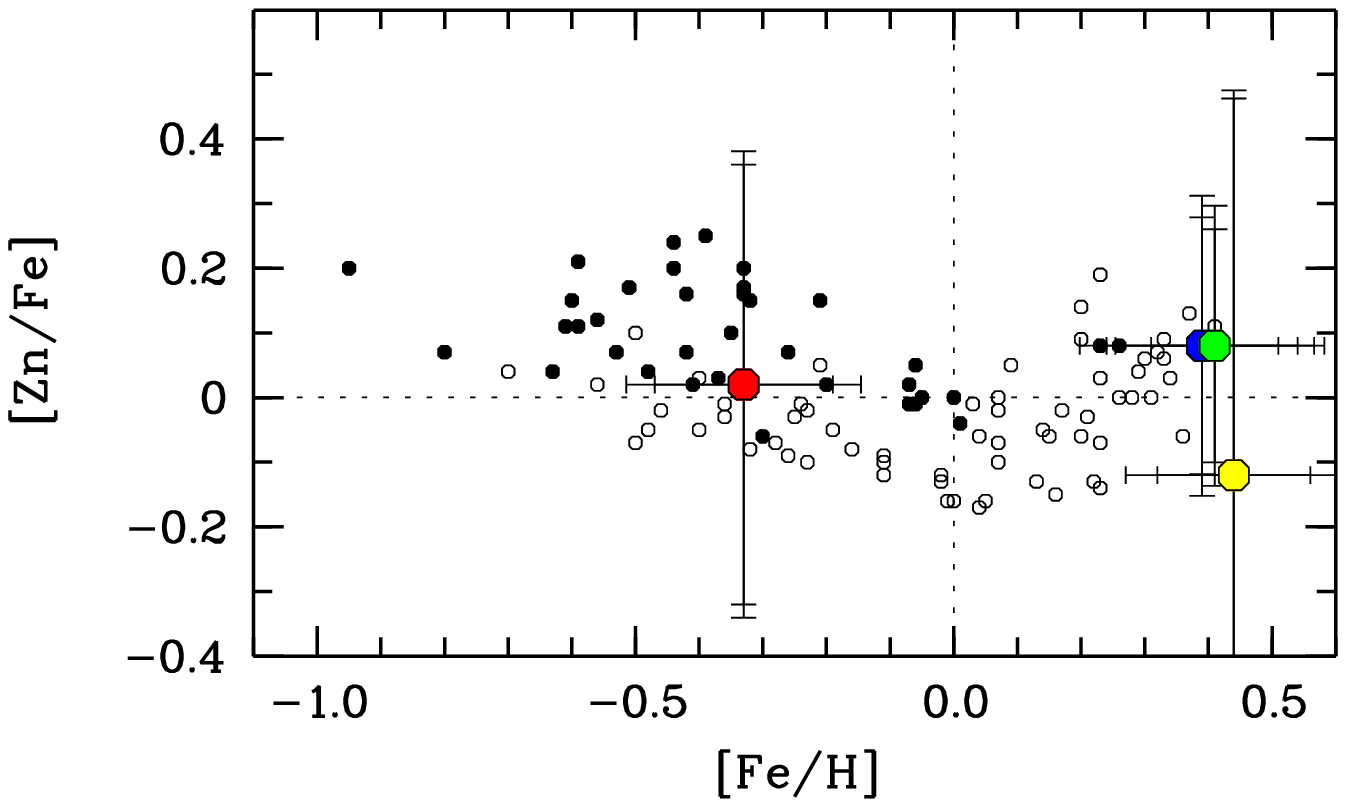}
                      \includegraphics[angle=-90,bb=315  40 510 440,clip]{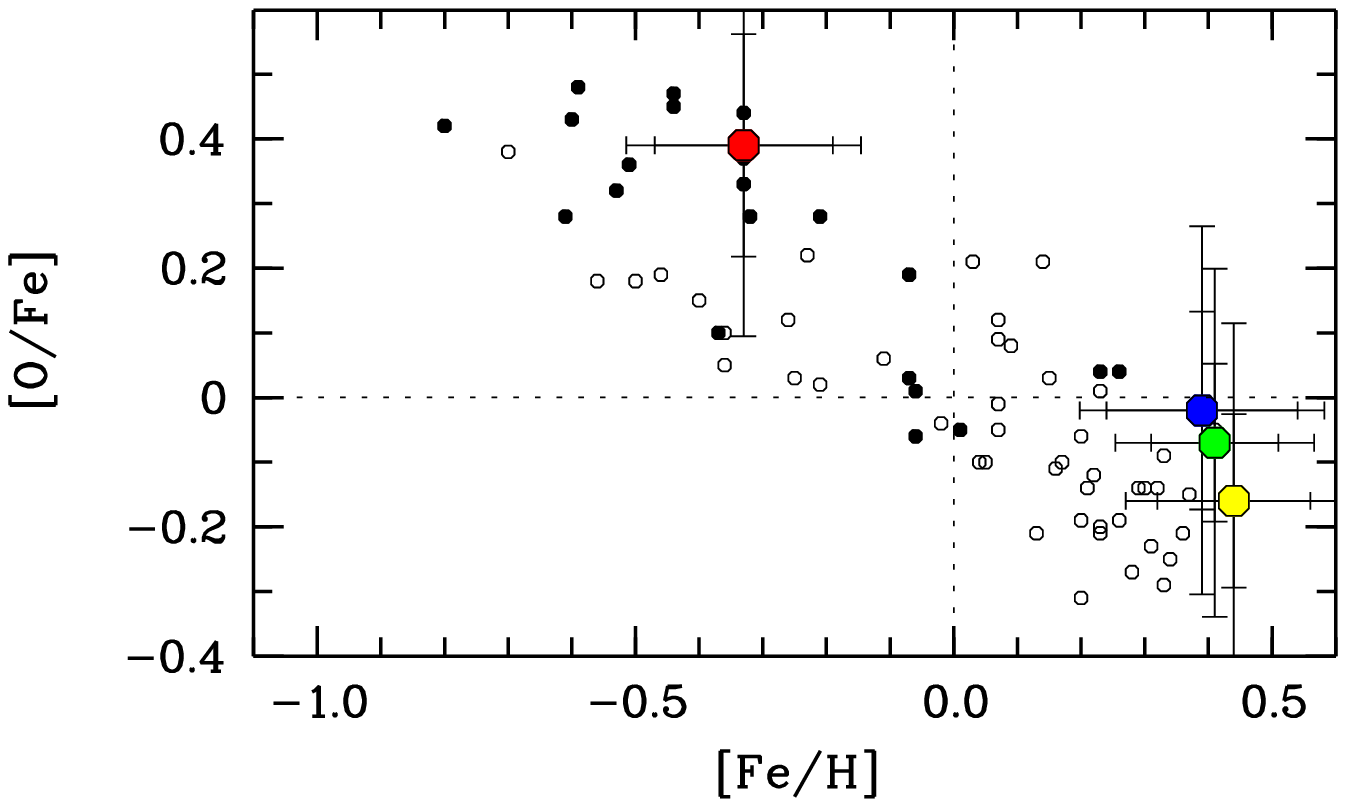}}
\resizebox{\hsize}{!}{\includegraphics[angle=-90,bb=325  40 555 430,clip]{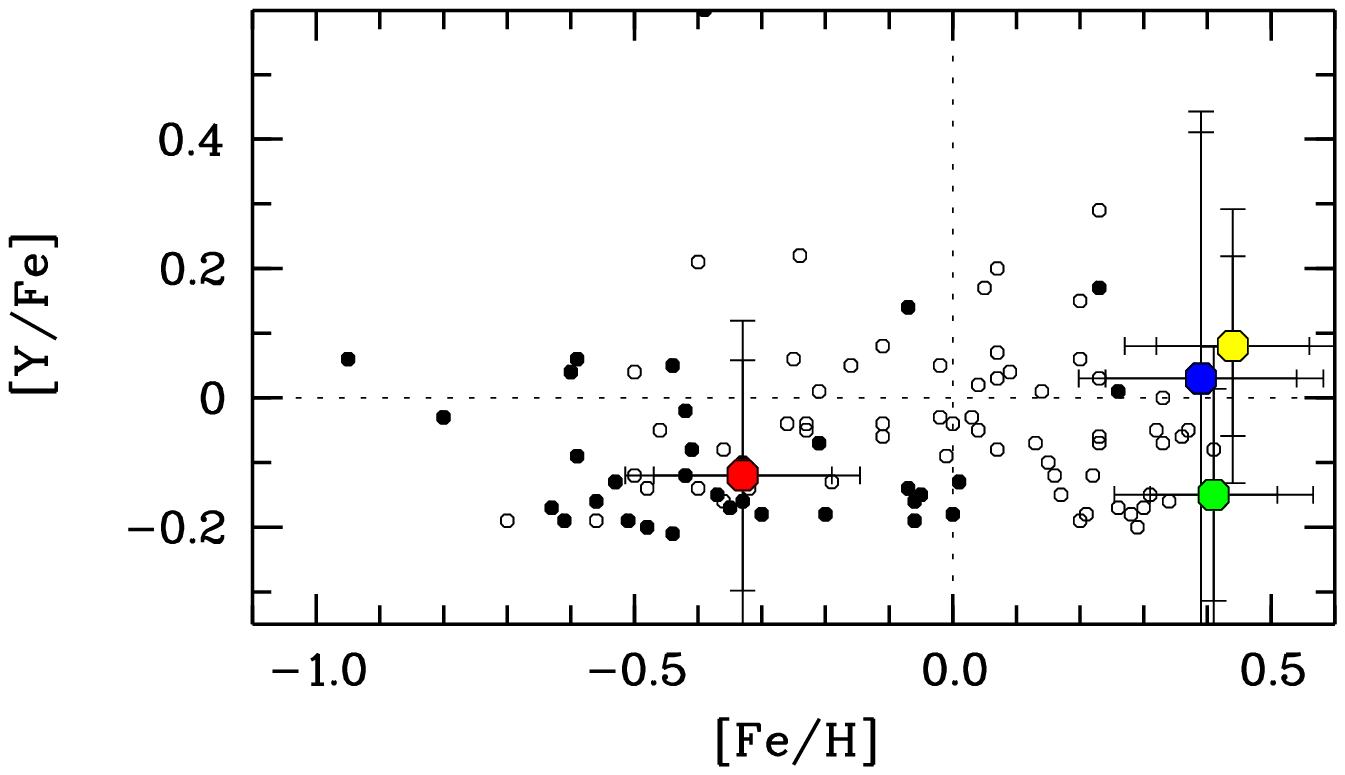}
                      \includegraphics[angle=-90,bb=325  40 555 440,clip]{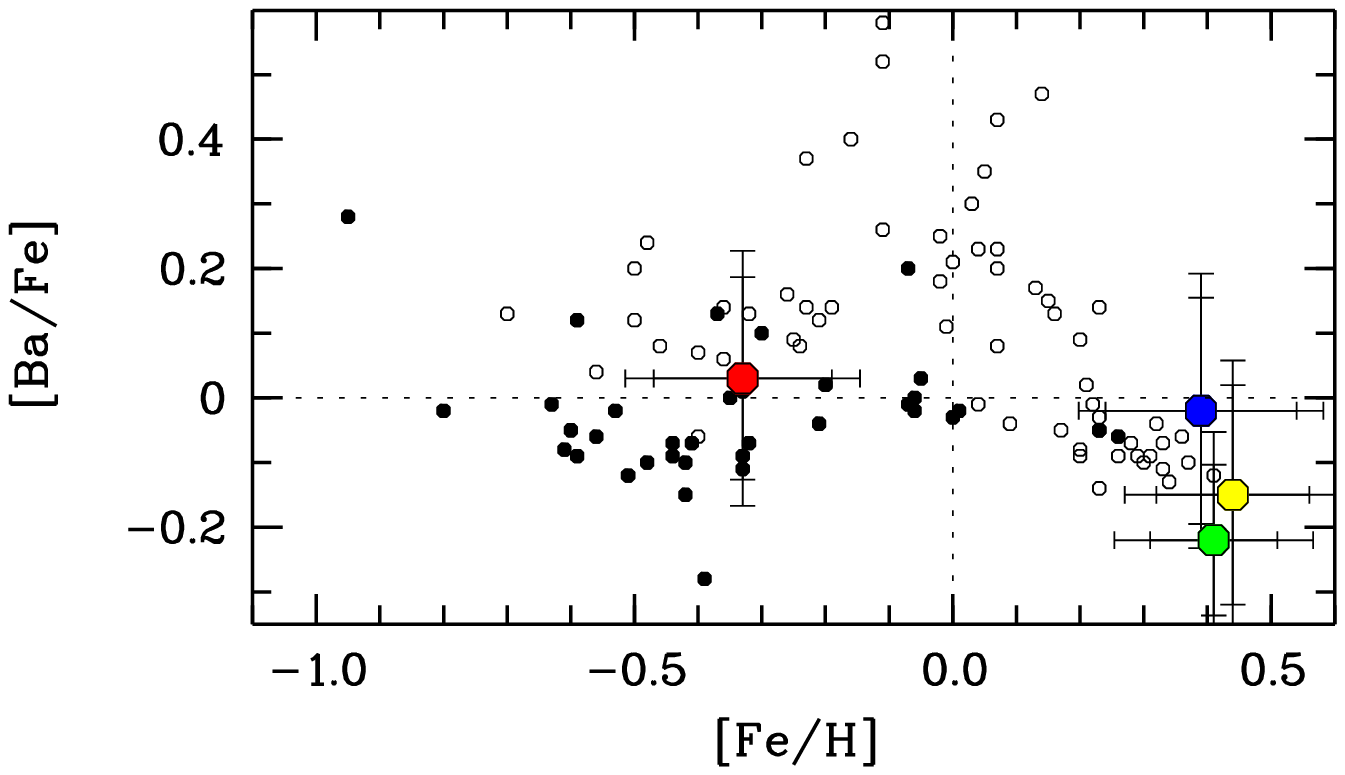}}
\caption{
        Abundance trends , [X/Fe] versus [Fe/H], for the thin and thick discs
        \citep[taken from][]{bensby2003,bensby2005}. Thin and thick disc
        stars are marked by open and filled circles, respectively.
        OGLE-2008-BLG-209S is marked by a red circle, OGLE-2006-BLG-265S 
        by a yellow circle, 
        MOA-2006-BLG-099S by a blue circle, and OGLE-2007-BLG-349S by a 
        green circle.
        The error bars have two cross bars. The inner one represent the
        standard deviation of the mean abundance, and the outer cross bar represent
        the total error, i.e. standard deviation of the mean abundance 
        and the uncertainty due to
        random errors in the stellar parameters, as given in 
        Table~\ref{tab:finalabundances}, added in quadrature.
        \label{fig:trends}}
\end{figure*}

\begin{figure*}
\resizebox{\hsize}{!}{\includegraphics[bb=175 28 550 590,clip,angle=-90]{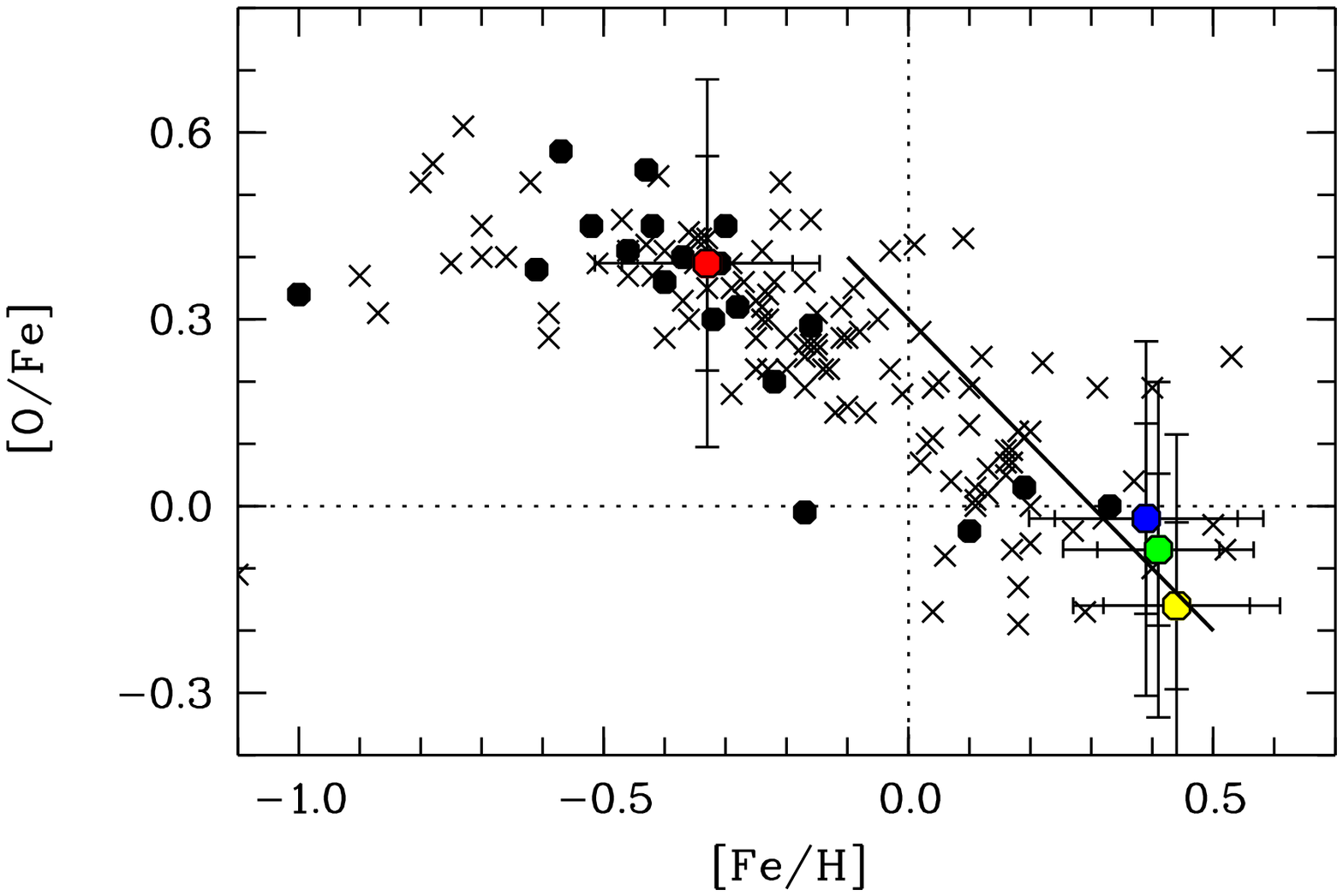}
                      \includegraphics[bb=175 28 550 590,clip,angle=-90]{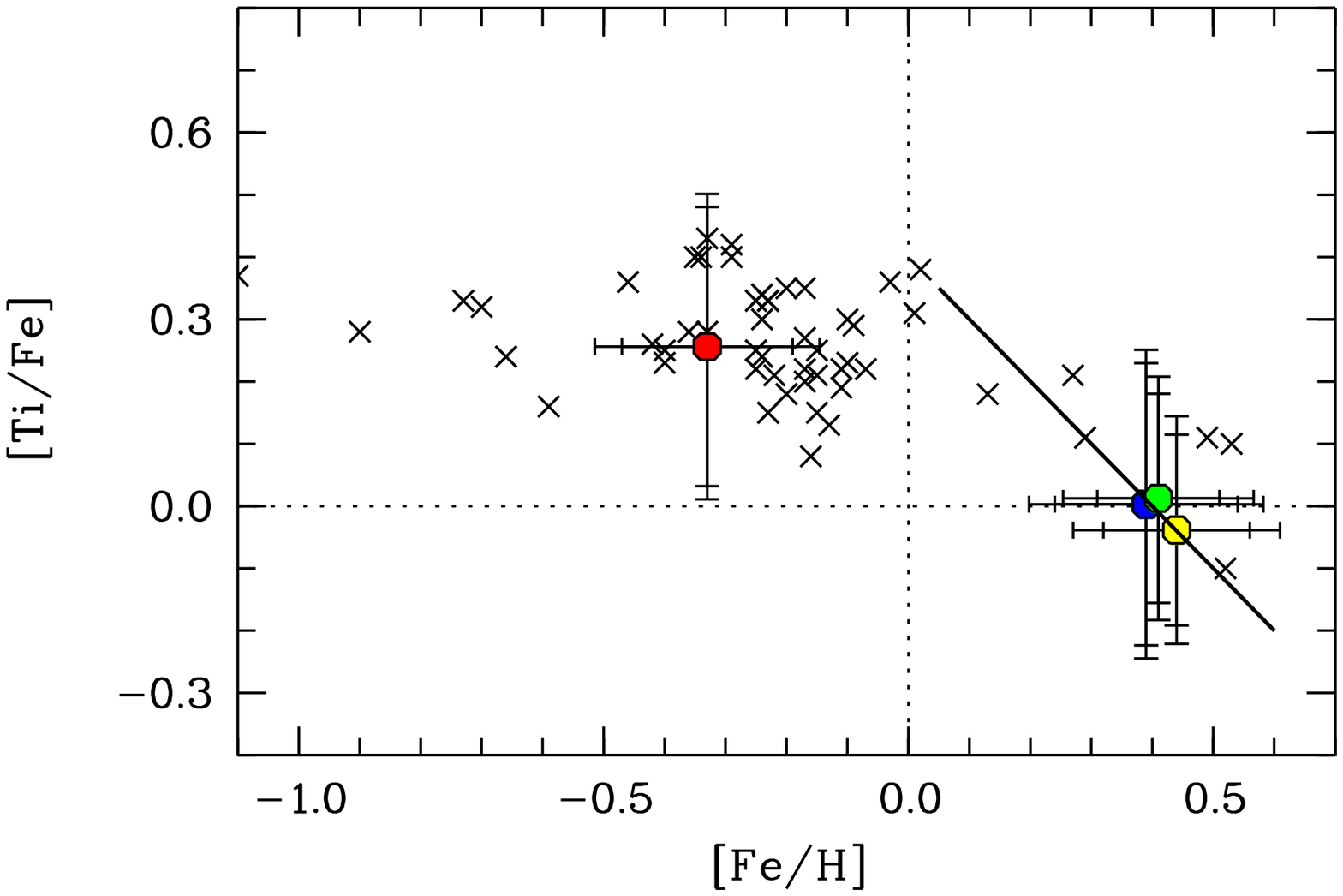}}
\resizebox{\hsize}{!}{\includegraphics[bb=175 28 550 590,clip,angle=-90]{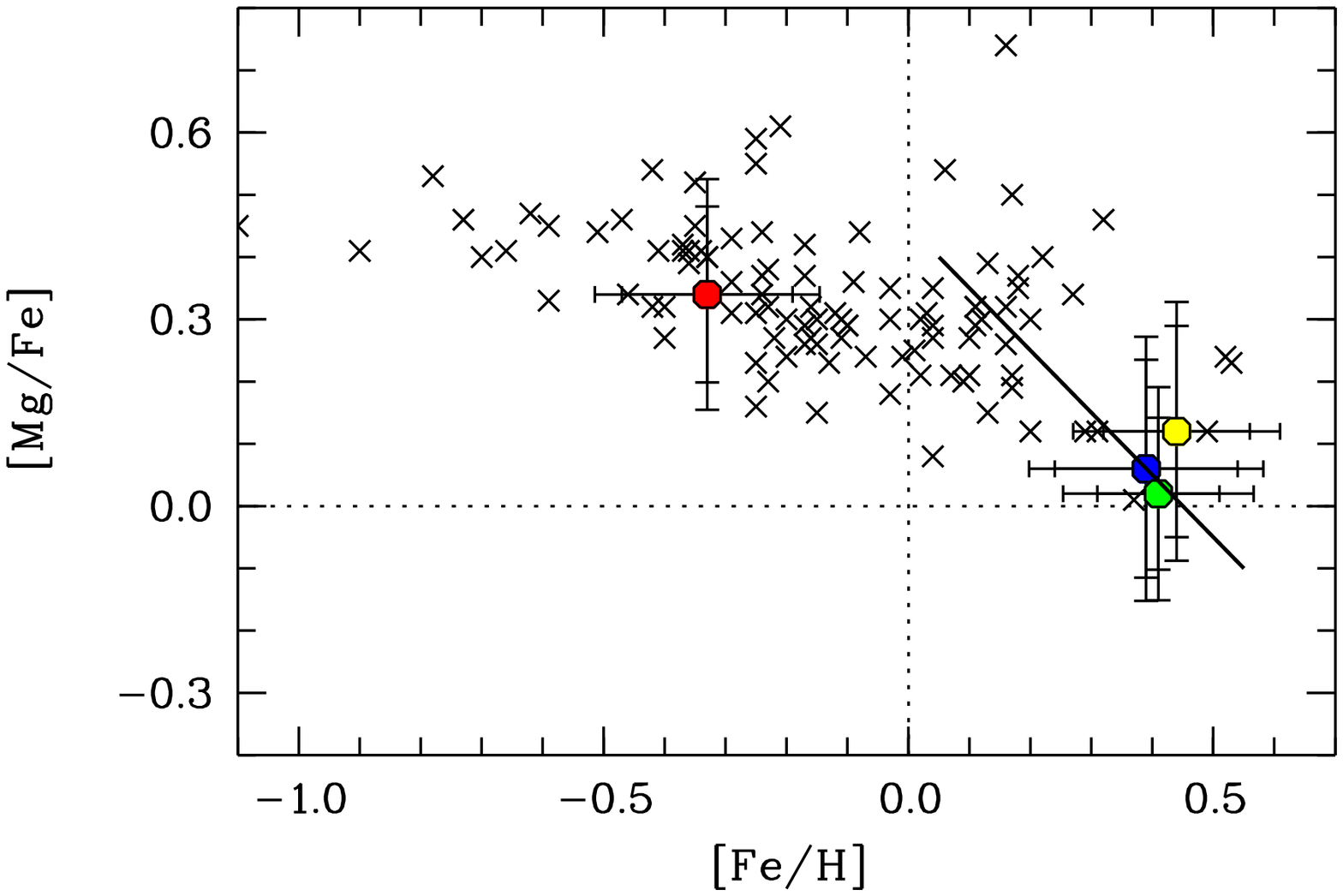}
                      \includegraphics[bb=175 28 550 590,clip,angle=-90]{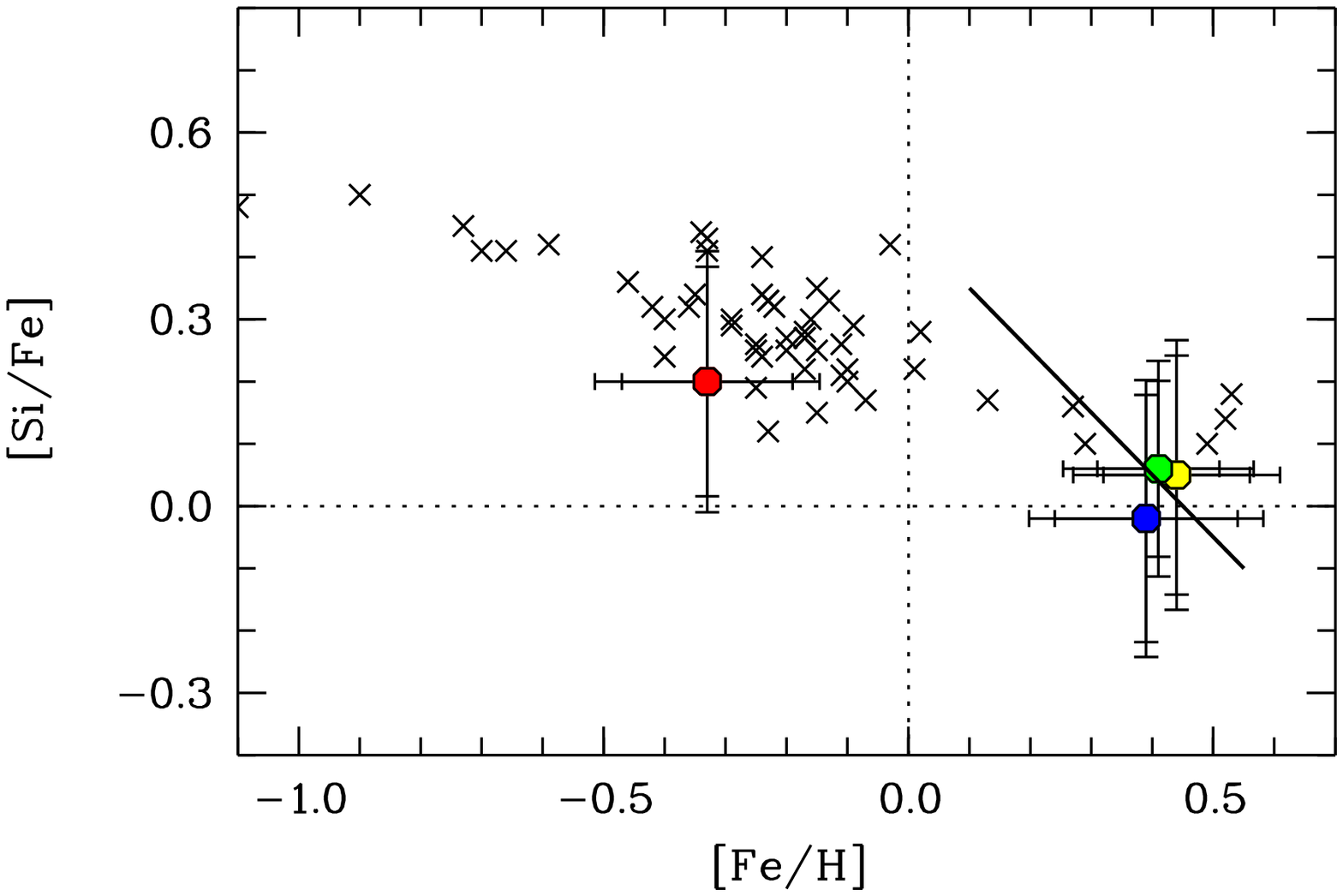}}
\caption{
Elemental abundance trends in the Bulge based on giant stars (marked by crosses) 
(O, Mg, Si, and Ti from \citealt{rich2005,rich2007,fulbright2007};
O and Mg from \citealt{lecureur2007}; O from \citealt{melendez2008}).
Nearby thick disc giant stars from
\cite{melendez2008} are marked by filled circles. The four stars in this study
are marked as in earlier figures.
The diagonal line marks the maximum slope that the [$X$/Fe] ratio can 
decrease in order to reach the levels of the three metal-rich dwarf stars
(the actual slope is however likely to be shallower, see discussion in
Sect.~\ref{sec:dwarfs}.)
        \label{fig:giants}}
\end{figure*}

\subsection{Elemental abundance trends in the Bulge}
\label{sec:dwarfs}

The Milky Way contains four major stellar populations; the halo,
the thin and thick discs, and the bulge. 

In Fig.~\ref{fig:trends} we show the thin and thick disc 
abundance trends from \citep{bensby2003,bensby2005}.
The stars of the thick disc have higher 
$\alpha$-to-iron abundance ratios than the stars of the 
thin disc. Specifically,
the stars of the thick disc show a constant high $\rm [\alpha/Fe]$
ratio for low metallicities and all the way up to $\rm [Fe/H]\approx-0.4$ 
after which a down-turn in $\rm [\alpha/Fe]$ occurs, steadily declining
towards solar values at solar metallicity. The stars of the thin disc, 
on the other hand, show a shallow decline in $\rm [\alpha/Fe]$ at lower 
metallicities that levels out at solar values at solar metallicities, 
where they merge with the thick disc trends. 

In Fig.~\ref{fig:trends} we also show the results for the four
dwarf/subgiant stars in the Bulge overplotted on the thin and thick 
disc abundance trends from \cite{bensby2003,bensby2005}. 
Generally, the abundance ratios for 
OGLE-2008-BLG-209S are similar to what is seen for the thick disc
at the same [Fe/H].
Interestingly, the metallicity of OGLE-2008-BLG-209S ($-0.33$\,dex)
is where the separation between the thin and thick discs is 
the greatest. 
If anything, this demonstrates that the observed [$\alpha$/Fe] ratios 
in OGLE-2008-BLG-209S make it
unlikely for it to be a thin disc star.

The three other Bulge dwarf stars, at super-solar
metallicities, show abundance ratios for the $\alpha$-elements
in close agreement with the metal-rich thin disc,
i.e., close to the solar values ($\rm [\alpha/Fe]\approx0$). 
Under the assumption that all four stars are all genuine members
of the Bulge, this means that the abundance trends of the Bulge, 
somewhere in between these points, should show a significant decrease in 
the $\alpha$-to-iron ratios, signalling the onset of chemical enrichment 
from low-mass stars.

Figure~\ref{fig:giants} shows the  abundance trends for O, Mg, Ti, and Si
in the Bulge as traced by K and M giant stars
\citep{rich2005,rich2007,lecureur2007,fulbright2007,melendez2008},
with the results for the microlensed dwarf and subgiant  stars overplotted. 
The main chemical characteristics of the Bulge using giant stars 
can be summarised as follows:
(1) The stellar formation history and chemical enrichment has been very fast. 
This is reflected by high [$\alpha$/Fe]
ratios even at solar metallicities; 
(2) The [O/Fe] ratio starts to decline
at roughly $\rm [Fe/H]\approx-0.3$, a signature of
the onset of chemical enrichment by low-mass stars.
(3) The [O/Fe]-[Fe/H] trends
in the Bulge and the thick disc are similar \citep{melendez2008}.
(4) Abundance trends based on other $\alpha$-elements such as Mg, Si, and Ti
are similar to the oxygen trend, with the exception that the
$\alpha$-enhancements for these elements, albeit slowly declining, 
tend to stay at somewhat higher levels with metallicity as compared to
oxygen \citep[see, e.g.][]{fulbright2007}. Also, there are significantly
fewer giant stars from these studies where abundances of Mg, and especially 
Ti and Si have been determined as compared to oxygen.
(5) Generally the abundance ratios from the giant stars
and the dwarf/subgiant stars seem to agree.

In order for a down-turn 
to occur in the first place in the [O/Fe]-[Fe/H] plot the 
oxygen enrichment (by massive stars) needs to decrease. 
The [O/Fe] ratio will then decrease, relative to Fe production, 
with [Fe/H] as a result 
of the continued Fe enrichment from low-mass stars. The maximum possible
decrease of the [O/Fe] trend  will occur if oxygen enrichment is 
shut off completely. The [O/Fe] ratio will then decrease by 
the same amount with which [Fe/H] increases. In the
[O/Fe]-[Fe/H] plot this correspond to a line with a slope 
equal to $-1$. For simplicity, lets assume
that this is the case. In Fig.~\ref{fig:giants} we show this
line, and it is clear that, in order for the [O/Fe] 
ratio path to be able to decline to the observed [O/Fe] ratio in
the three metal-rich Bulge dwarf stars, the oxygen production needs
to be abruptly shut off at a metallicity not higher 
than $\rm [Fe/H]\approx -0.1$. It is of course unlikely
that the oxygen enrichment should turn off ``over-night", instead
it should be a gradual decrease, i.e. the slope of the line should 
be less than what we have indicated in Fig.~\ref{fig:giants}. 
If so, it seems likely that the
turn-over in the [O/Fe]-[Fe/H] trend as traced by dwarf and subgiant stars
happens at approximately the same metallicity as for the giant stars,
i.e. at $\rm [Fe/H]\approx -0.3$ to $-0.4$ \citep{melendez2008}. 
This also happens to
be where we see the ``knee" in the thick disc [$\alpha$/Fe]-[Fe/H]
trends \citep{bensby2003,bensby2004,bensby2005}.
Similar lines have been drawn in the Mg, Si, and Ti plots in Fig.~\ref{fig:giants},
indicating that the enrichment from low-mass stars started at $\rm [Fe/H]\approx0$
(under the assumption that the enrichment of these elements from massive stars
turned off abruptly).

\subsection{A comparison of Bulge and inner disc dwarf stars}

\begin{figure}
\resizebox{\hsize}{!}{
\includegraphics[angle=-90,bb=100 28 360 590,clip]{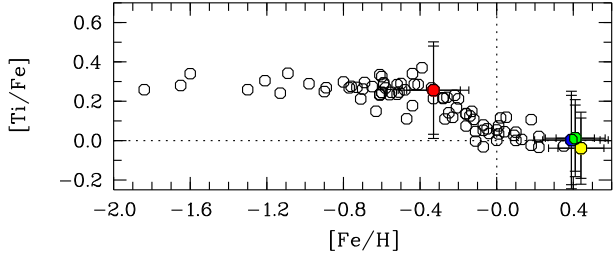}}
\caption{ 
82 nearby dwarf stars, selected from Bensby et al.~(2009, in prep.),
that have kinematic properties making them likely to come from the 
inner Galactic disc/Bulge. The kinematic criteria are the same as
in \cite{pompeia2002a}, i.e., 
$e>0.25$, $Z_{\rm max}<1.0$\,kpc, and  $R_{\rm min}<4$\,kpc.
OGLE-2008-BLG-209S (red circle), OGLE-2006-BLG-265S (yellow circle), 
and MOA-2006-BLG-099S (blue circle) 
have also been included.
\label{fig:bulgelike}
}
\end{figure}

There is currently no
in situ stellar sample that traces the inner disc regions of the
Galaxy using detailed elemental abundances. A possible way to form
a compatible sample would be to sample high-velocity stars with 
orbits that make them likely to have
come from the inner parts of the Galactic disc.
For instance, \cite{pompeia2002a} studied a sample of, so called,
nearby ``bulge-like" dwarf stars. These were defined as stars with 
highly eccentric orbits ($e>0.25$),
that do not reach more than 1\,kpc from the Galactic plane 
($Z_{\rm max}<1$\,kpc), and that have a maximum peri-galactic 
distance ($R_{\rm p}$) of 3 to 4\,kpc. 
In Bensby et al.~(2009, in prep) we have $\sim700$ stars and
82 of those stars fulfil the ``bulge-like" criteria by
\cite{pompeia2002a}. Figure~\ref{fig:bulgelike}
shows the [Ti/Fe]-[Fe/H] trends for these stars
together with the four microlensed Bulge dwarf/subgiant stars.
The observed trend for the ``bulge-like"  disc sample 
is very well defined with 
very little scatter. At sub-solar [Fe/H] there are no signs
of [Ti/Fe] ratios seen  in the local thin disc. This might not 
come as an
surprise as we are picking stars on highly eccentric orbits, and if we were
to trace the thin disc trend closer to the Galactic centre, those stars
would likely be on on almost circular orbits and hence never cross the 
Solar orbit. 
It could also be that these disc stars on highly
eccentric orbits belong to an inner disc population, distinct
from the local thin and thick discs. This will be investigated further
in Bensby et al.~(2009, in prep.). 

Plotting the microlensed Bulge dwarf and subgiant stars on top of the
``bulge-like" disc sample (using the definition from \citealt{pompeia2002a}) 
in Fig.~\ref{fig:bulgelike} 
we see that OGLE-2008-BLG-209S nicely fits into the trend
and that the other three Bulge dwarf stars form a 
metal-rich extension of the trend.

\subsection{The metallicity distribution of the Bulge}

In Fig.~\ref{fig:zoccali} we plot the metallicity distribution of the
521 giant stars in the Bulge observed by \cite{zoccali2008}
in Baade's window $(l,b)=(1.14,-4.18)$, and two other Bulge fields
at $(l,b)=(0.21,-6.02)$ and $(l,b)=(0.0,-12.0)$.
Also shown is the MDF of the 24 giant stars analysed by
\cite{fulbright2007}, on which we have overplotted
the metallicities of the four Bulge stars analysed
in this work. Note that the distribution of the \cite{fulbright2007}
is {\sl not} representative of the MDF in the Bulge.  Their sample
is designed to probe the full range of [Fe/H], not to be a 
representative sample.
As can be seen the Bulge MDF based on the giant stars range from
very low metallicities, $\rm [Fe/H]\approx-1.3$, to super-solar 
values comparable to the three previously 
published dwarf stars at $\rm [Fe/H]\approx0.4$--0.5 
\citep{johnson2007,johnson2008,cohen2008}.
The average metallicity is $\rm [Fe/H]\approx-0.2$ \citep{zoccali2008}.

An issues that has been discussed against results 
based on dwarf stars that have been microlensed 
is that the microlensing is affecting the 
spectrum of the source star and therefore
we are not getting the right answer. This is why these stars 
(for some reason) are all very metal-rich \citep{zoccali2008}.
OGLE-2008-BLG-209S, the most recently analysed microlensed star, although
being a subgiant, has
a metallicity of $\rm [Fe/H]=-0.33$,
and proves that it is indeed possible to find metal-poor microlensed 
stars. Also, since OGLE-2008-BLG-209S is approximately at the
distance of the stars in Baade's Window (because it's very close 
spatially to BW, see Fig.~\ref{fig:allevents}), its metallicity is
very compatible with the MDF of giant stars
if they are spatially coincident (which is the case of OGLE-2008-BLG-209S and
the giant stars by \citealt{zoccali2008}). So this
strengthens the idea that the analysis of microlensed dwarf and subgiant stars 
needs to be taken seriously.

\subsection{Formation and chemical history of the Bulge}
\label{sec:bulgehistory}

There are essentially three main formation scenarios for how 
galactic bulges may form. 

Firstly,  we have the classical scenario 
where the bulge form in the final phases of the monolithic collapse of the 
protogalactic cloud \citep{eggen1962}. 
First the halo forms and as the collapse progresses
also the bulge. The similarity between the distributions of angular 
momentum in the Bulge and halo suggest that this could 
be the case \citep{wyse1992}. 
Secondly, it is possible that bulges
form from mergers in a $\Lambda$CDM Universe. 

\begin{figure}
\resizebox{\hsize}{!}{
\includegraphics[angle=-90,bb=20 60 560 580,clip]{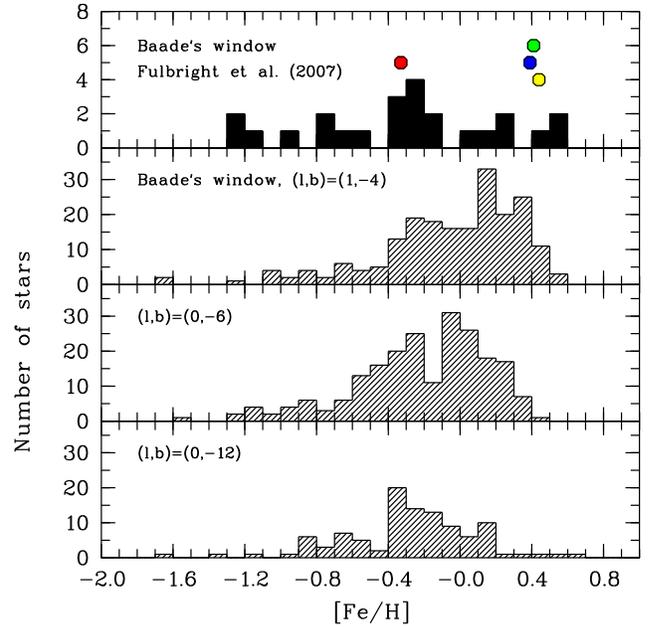}}
\caption{Metallicity distribution of the Bulge. Top panel shows
metallicity histogram of the 24 giant stars from \cite{fulbright2007},
on which the metallicities of the four Bulge dwarf/subgiant 
stars analysed in this work,
marked as in previous plots (arbitrarily shifted vertically).
The three bottom panels show the MDF of the giant stars in three different
Bulge fields from \cite{zoccali2008}.  The $(l,b)=(1.14,-4.18)$ field
has 244 stars, the $(l,b)=(0.21,-6.02)$ field 213 stars, and the
$(l,b)=(0.0,-12.0)$ field 104 stars.
\label{fig:zoccali}}
\end{figure}

The outcome from these two scenarios is that the abundance trends 
in the Bulge  could show many types of signatures.
Depending on the star formation history, it is possible to get
high $\alpha$-to-iron ratios extending to super-solar metallicities. 
However, as both of these scenarios happen in the
very early phases of the formation history of the Galaxy,
they would result in a Bulge that contain very old stars, 
as old as those found in the stellar halo. Even if not
exclusively so, old stars are generally what is found
in the Bulge \citep[e.g.][]{feltzing2000b,ortolani1995}. 
However, in the collapse model, the average metallicity 
should increase as the collapse progresses inwards, resulting 
in metallicity gradients.
Tentative evidence for a small metallicity gradient in the
Bulge was found by \cite{zoccali2008}, see Fig.~\ref{fig:zoccali}.
For the hierarchical merger scenario in a $\Lambda$CDM Universe,
where the in-falling ``pieces" will be randomly distributed within 
the Bulge, no metallicity gradients are expected. 

The third scenario is that bulges form through secular evolution, 
in which a bulge-like  structure is
built up through the slow rearrangement of energy and mass as a result of 
interactions between for instance a galactic bar and a galactic disc.
If secular evolution dominates then it is
quite likely that  abundance trends in the bulge would mimic those of the
central galactic disc as that is where the material and energy would
come from to form the bulge. 
For the Milky Way bulge,
the secular scenario has until recently been seen as quite unlikely
because of the differing abundance patterns in the Bulge to those
in the disc \citep[e.g.][]{fulbright2007}. 
However, recently \cite{melendez2008}
carried out a truly differential study of Bulge and thin and thick
disc giant stars and found that they are indeed very similar;
the Bulge almost perfectly mimics the oxygen abundance trends
of the nearby thick disc. Our results do not contradict this similarity.

As also pointed out by \cite{melendez2008},
the similarity of the Bulge and the nearby thick disc oxygen trends, 
even though these two population clearly are spatially separated within the
Galaxy, indicates that they have had very similar, but not 
necessarily the same, chemical enrichment histories.
Observations of distant galaxies suggest that the secular scenario 
is very common \citep{genzel2008}, and with the new abundance
results we find that it is also a plausible
scenario for the Milky Way bulge. 

\section{Summary}
\label{sec:summary}

We have performed a detailed elemental abundance analysis of 
OGLE-2008-BLG-209S, a microlensed subgiant  star in the Bulge.
In contrast to the three previously published
microlensed Bulge dwarf stars, that all turned out to be
extremely metal-rich, having metallicities greater
than $\rm [Fe/H]>0.35$ 
\citep{johnson2007,johnson2008,cohen2008},
OGLE-2008-BLG-209S has a more modest sub-solar metallicity of
$\rm [Fe/H]\approx -0.33$.
Interestingly, the metallicity of OGLE-2008-BLG-209S
is where the separation in the [$\alpha$/Fe] ratio 
between the thin and thick disc stars is the greatest.
Furthermore, the abundance pattern of OGLE-2008-BLG-209S is similar 
to what is found in nearby thick disc dwarf stars and what is found
in giant stars in the Bulge, i.e. enhanced $\alpha$-to-iron ratios 
(e.g., O, Mg, Si, Ca, Ti).

We have also carried out a re-analysis of the three  
microlensed dwarf stars, previously published by 
\cite{johnson2007,johnson2008} and \cite{cohen2008}.
This was done in order to get a homogeneous
sample of Bulge dwarf and subgiant stars analysed with the same methods as
for the large sample of nearby thin and thick disc
dwarf stars by \cite{bensby2003,bensby2005}.
With some exceptions we see good agreement 
in abundances between our results for MOA-2006-BLG-099S, 
OGLE-2006-BLG-265S, OGLE-2007-BLG-349S and those in 
\cite{johnson2007,johnson2008,cohen2008}.
For instance, the metallicity of OGLE-2006-BLG-265S is lower 
in our new analysis ($\rm [Fe/H]=+0.44$) than what
\cite{johnson2007} found ($\rm [Fe/H]=+0.56$).

The three stars for which we could estimate ages are (using method 1)
8.5\,Gyr, 7.5\,Gyr, and 13\,Gyr old, respectively. 
These ages are 
higher than what we see for thin disc  stars at these metallicities.
However, the stars of the thick disc
are found to be older than the stars of the thin disc.
This could suggest that the metal-rich stars
of the Bulge have ages comparable to what we see in the
metal-rich thick disc (at solar metallicities, \citealt{bensby2007letter2}). 
It should be cautioned that individual
ages based on isochrones could be subject to substantial systematic
and random uncertainties. Therefore, ages for many more bulge 
dwarf and subgiant stars must be determined before firm conclusions can be drawn. 

We have also considered several methods in analysing the stars
and find that elemental abundance ratios based on
the methods are in good agreement, showing small or negligible 
differences. We would like to stress that the differences 
between the methods we used to get abundances from Bulge dwarf stars
have been quantified, and that our investigation shows that 
one always should be very careful in combining information from differnet 
studies unless all "facts" are clearly explained (such as the solar abundances 
used for normalisation, etc.).

The above results are based on less than
a handful of dwarf/subgiant stars. It is very important to further
confirm these results with more observations. However, 
high-magnification microlensing
events, that currently seem to be the only feasible way of getting
good quality high-resolution data of Bulge dwarf stars, are very rare.
Out of the yearly thousands of detected microlensing events from the
OGLE monitoring of the Bulge, only a dozen are dwarf events. 
As it is impossible to know beforehand
when these will occur, it is necessary to have a flexible
observing schedule to be able to catch them. As such, the
Rapid Response Mode with the UVES spectrograph
on the Very Large Telescope should be optimal.

\begin{acknowledgement}

 We would like to thank Bengt Gustafsson, Martin Asplund, 
 Bengt Edvardsson, and Kjell
 Eriksson for usage of the MARCS model atmosphere program and their
 suite of stellar abundance (EQWIDTH) programs.  
 Paul Barklem is also thanked for helping us to enlarge the
 grid of stellar model atmospheres.
 S.F. is a Royal Swedish Academy of Sciences
 Research Fellow supported by a grant from the Knut and Alice
 Wallenberg Foundation. Work by A.G. was supported by
 NSF Grant AST-0757888. J.C. and W.H. are grateful to NSF grant 
 AST-0507219  for partial support.
 A.U. acknowledges  support by the Polish MNiSW grant N20303032/4275.
 J.S. is supported by a Marie Curie Incoming International Fellowship.
 D.A. and S.F. thank the Swedish Research Council for a dedicated
 travel grant that enabled D.A. to travel to Las Campanas for this 
 observing run.
\end{acknowledgement}
\bibliographystyle{aa}
\bibliography{referenser}


\end{document}